\documentclass[preprint,10pt]{elsarticle}

\usepackage{amssymb}

\usepackage[nodots,nocompress]{numcompress}

\biboptions{square}

\journal{Nuclear Physics B}

\begin{document}

\begin{frontmatter}



\title{Action Principles for Transgression and Chern-Simons AdS Gravities}


\author{Pablo Mora}
\ead{pablomora@cure.edu.uy}
\address{Centro Universitario Regional Este (CURE), Universidad de la Rep\'ublica, Uruguay, Ruta 9 Km 207, Rocha, Uruguay}

\begin{abstract}
Chern-Simons gravities are theories with a lagrangian given by a Chern-Simons form constructed from a space-time gauge group. In previous investigations  we showed that, for some special field configurations that are solutions of the field equations, the  extension from Chern-Simons to Transgression forms as lagrangians, motivated by gauge invariance, automatically yields the boundary terms required to regularize the theory, giving finite conserved charges and black hole thermodynamics.

Further work by other researchers showed that one of the action functionals considered in the above mentioned work yields a well
defined action principle in the metric (zero torsion) case and for asymptotically Anti de Sitter (AdS) space-times. 

In the present work we consider several action functionals for Chern-Simons AdS gravity constructed from Transgression forms,
and show the action principles to be well defined and the Noether charges and Euclidean action to be finite for field configurations satisfying only that the gauge field curvature (field strength) for the AdS gauge group is asymptotically finite.

For that purpose we consider an asymptotic expansion of the vielbein and spin connection that may be regarded as a perturbation
of an AdS space-time, but allowing a non zero torsion.

Our results are of potential interest for Lovelock gravity theories, as it has been shown that the boundary terms
dictated by the transgressions for Chern-Simons gravities are also suitable to regularize Lovelock theories.  
\end{abstract}


\end{frontmatter}















\section{Introduction}

Chern-Simons (CS) gravities in 2+1 dimensions were introduced and studied in Ref.\cite{chern-simons-2+1-1,chern-simons-2+1-2}, extended to higher dimensions by Chamseddine in Refs.\cite{chamseddine-1,chamseddine-2} and to the supersymmetric case in Refs.\cite{banados-troncoso-zanelli-1,banados-troncoso-zanelli-2}. These theories have been further studied and extended in several aspects aspects in many works, having very interesting properties from the point of view of their dynamic and symmetries, as they are true gauge theories of gravity with solutions that correspond to black holes, black branes as well as other solutions. For a recent review of this topic with an extensive and comprehensive list of references see \cite{zanelli-lectures} (for older reviews see \cite{troncoso-zanelli}). Chern-Simons AdS gravities, which are the subject of the present article, are Chern-Simons gauge theories with tha Anti de Sitter (AdS) group as their gauge group.

Chern-Simons forms are not strictly invariant under gauge transformations, but only quasi-invariant, meaning that they change by a closed form. Transgression forms (see for instance \cite{nakahara, alvarez}) are extensions of Chern-Simons forms that are strictly gauge invariant, but are functionals of two gauge fields $A$ and $\overline{A}$, unlike CS forms wich depend only on one gauge field $A$. Transgressions have been considered as actions for physical theories in refs.\cite{potsdam,transgression-branes-1,transgression-branes-2,more-transgressions-1,more-transgressions-2,IRS-1,IRS-2,sarda-1,sarda-2,tesis,motz1,motz3}, where several aspects of this models have been explored. In particular in refs.\cite{motz1,motz3} it was shown that the extensions of Chern-Simons AdS gravities dictated by the transgressions have the built-in boundary terms that regularize the action, in the sense of giving a finite action, finite Noether conserved charges and the right black hole thermodynamics, unlike what happens in CS theories, where those quantities are infinite unless one regularizes them by hand. The results of refs.\cite{motz1,motz3}
may be regarded as strong but circumstantial evidence, as only special field configurations were considered.

The work of refs.\cite{Olea-Miskovic-1,Olea-Miskovic-2} goes further, showing that one of the action functionals considered in the above mentioned work yields a well
defined action principle in the metric (zero torsion) case and for asymptotically Anti de Sitter (AdS) space-times. 

Here we will discuss  possible action functionals for Chern-Simons AdS gravity constructed from Transgression forms, along the lines of refs.\cite{motz1,motz3}. We will prove that the action principles are well defined and that the Noether charges and Euclidean action are finite for field configurations satisfying only that the gauge field curvature (field strength for the AdS gauge group) is asymptotically finite.
In order to impose  the finite asymptotic gauge curvature condition we need to consider an asymptotic expansion of the vielbein and spin connection that may be seen as a perturbation of AdS space-time, in the same sense that the Fefferman-Graham expansion
\cite{fefferman-graham}, but allowing a non vanishing torsion. The finite curvature condition allow us to determine the leading order of the asymptotic behavior of the relevant fields, which is just what we need.

The structure of the present article is the following:

In Section 2  we review Chern-Simons and Transgression forms. Section 3 reviews Transgressions as actions for physical
theories. Section 4 review Transgression and Chern-Simons AdS gravity.
Section 5 is devoted to determine the asymptotic dependence of the fields required to have a finite gauge curvature asymptotically. Section 6 deals with action principles of the kind that we call "Backgrounds", as one may seen those actions as regulated by a sort of background subtraction \cite{motz1,motz3}. In that context we distinguish two possibilities: two dynamical field configurations or a dynamical field configuration and 
a non dynamical field configuration (which we will call "AdS vacuum"). In Section 7 we consider another possible action principle,
which we call "Kounterterms action principle"\cite{motz1,motz3,Olea-Miskovic-1,Olea-Miskovic-2}, that can be understood as coming from a Transgression lagrangian with a dynamical field configuration and non dynamical
one  (which we will call "Kounterterms vacuum"). In Section 8 some known solutions are recast in the 
standard for of the coordinates used in Section 5, and conserved charges are 
computed in the framework of the previous sections.

\section{Chern-Simons and Transgression forms}

\subsection{Transgressions}

Chern-Simons forms\footnote{For the details of the mathematics of Chern-Simons and Transgression forms and references see \cite{nakahara}.} $\mathcal{C}_{2n+1}(A)$ are differential forms defined for
a connection $A$, which under gauge transformations of that connection transform by a closed form, so are say to be \textit{quasi invariant}. Transgression forms $\mathcal{T}_{2n+1}$ are a generalization of Chern-Simons forms that depend on two gauge connections $A$ and $\overline{A}$ and are strictly gauge invariant if both connections are subjected to the same gauge transformation.  The use of this forms as lagrangians for physical theories, or as a template for actions for physical theories was discussed in references \cite{motz1, motz3}. Transgressions can be written (see e.g., \cite{nakahara}) as the difference of two Chern-Simons forms plus an exact form 
\begin{equation}
\mathcal{T}_{2n+1}(A,\overline{A})=\mathcal{C}_{2n+1}(A)-\mathcal{C}_{2n+1}(\overline
{A})-d\mathcal{B}_{2n}\left(  A,\overline{A}\right)\label{transgression-1}
\end{equation}
where $\mathcal{T}_{2n+1}(A,\overline{A}=0)=\mathcal{C}_{2n+1}(A)$,  or explicitly as 
\begin{equation}\label{transgression}
\mathcal{T}_{2n+1}\left(  A,\overline{A}\right)  =(n+1)\int_{0}^{1}
dt\ <\Delta{A}F_{t}^{n}>\label{transgression-2}
\end{equation}
where\footnote{Here wedge product between forms is assumed.} 
$A_{t} = tA+(1-t)\overline{A}=\overline{A}+t\Delta A $
is a connection that interpolates between the two independent gauge potentials
$A$ and $\overline{A}$. The Lie algebra-valued one-forms\footnote{Notation: In what follows upper case Latin indices from the beginning of the alphabet $A,~B,~C,...$ are space-time indices with values from $0$ to $d-1=2n$; upper case Latin indices from the middle of the alphabet $I,~J,~K,...$ are space-time indices with values from $0$ to $d-1=2n$ but different from 1 (with 1 corresponding to a "radial" coordinate, or a coordinate along the direction normal to the boundary); lower case Latin indices from the beginning of the alphabet $a,~b,~c,...$ are tangent space (or Lorentz) indices with values from $0$ to $d-1=2n$; lower case Latin indices from the middle of the alphabet $i,~j,~k,...$ are tangent space (or Lorentz) indices with values from $0$ to $d-1=2n$ but different from 1 (with 1 identified to a "radial" direction, or a direction normal to the boundary in tangent space). The index $\alpha$ labels the generators $G_{\alpha}$ of the Lie group considered and takes values from 1 to the dimension of the group.} $A=A_{A}^{\alpha}G_{\alpha}\ dx^{A}$ and $\overline{A}=\overline{A}_{A}^{\alpha}G_{\alpha}\ dx^{A}$ are
connections under gauge transformations, $G_{\alpha}$ are the generators of the gauge group $\mathcal{G}$ (elements of its Lie algebra $\mathfrak{G}$) and
$<\cdots>$ stands for a symmetrized invariant trace in the Lie algebra (or equivalently for the contraction with a symmetric invariant tensor of the group). 
The corresponding curvature is
$F_{t}=dA_{t}+A_{t}^{2}=t F +(1-t)\overline{F}-t(1-t)(\Delta A)^{2}$.  Setting $\overline{A} =0$ 
in the transgression form yields the Chern-Simons form for $A$.
If $g$ is an element of $\mathcal{G}$, then a gauge transformation of $A$ is given by $A^g=g^{-1}[A+d]g$ and the field strength transforms covariantly as $F^g=g^{-1}Fg$. If $\overline{A}$ is transformed with the same group element, then $\Delta A$ and $F_t$ transform covariantly, and from eq.[~\ref{transgression}] it is clear that the transgression is gauge invariant in that case. The case where $A$ is transformed but $\overline{A}$ is not is considered in the next subsection, and it is relevant to compute gauge anomalies with backgrounds.
  
We will use that the variation of the transgression under infinitesimal variations of $A$ and $\overline{A}$ is
\begin{eqnarray}
\delta\mathcal{T}_{2n+1}=(n+1) <F^n\delta A>-(n+1)<\overline{F}^n\delta \overline{A}>\nonumber\\
-n(n+1)d\{\int _0^1dt<\Delta AF_t^{n-1}\delta A_t>\}\label{variation-transgression}
\end{eqnarray} 
 
\section{Transgression forms and Actions}

Transgression forms have been used to define actions $I_{Trans}$ for physical theories \cite{motz1,motz2,motz3} through
\begin{equation}
I_{Trans}=\int _{\cal M}\mathcal{C}_{2n+1}(A)-
\int _{\overline{\cal M}}\mathcal{C}_{2n+1}(\overline
{A})-\int _{\partial {\cal M}}\mathcal{B}_{2n}\left(  A,\overline{A}\right)\label{transgression-action}
\end{equation}
Notice that each Chern-Simons form is integrated in a different bulk manifold, but $\partial {\cal M}=\partial \overline{{\cal M}}$ is the common boundary of both manifolds\footnote{This generalization was motivated by the application to Chern-Simons AdS gravity theories, for which the manifolds ${\cal M}$ and $\overline{\cal M}$
may even have different topologies (for instance black hole and AdS space-times).}. 

Also observe 
that eq.(\ref{transgression-action}) is not just the result of integrating eq.(\ref{transgression-1}), $I_{Trans}\neq \int _{\cal M}\mathcal{T}_{2n+1}(A,\overline{A})$, as there 
are two different bulk manifolds involved. 

It is however important that
\begin{eqnarray}
\delta I_{Trans}=(n+1)\int _{\cal M} <F^n\delta A>-(n+1)\int _{\overline{\cal M}}<\overline{F}^n\delta \overline{A}>-\nonumber\\
-n(n+1)\int _{\partial {\cal M}}\{\int _0^1dt<\Delta AF_t^{n-1}\delta A_t>\}\label{variation-trans-action}
\end{eqnarray}
We will need the previous equation in what follows, and it implies in particular that $I_{Trans}$ is gauge invariant with the AdS gauge group
\footnote{That eq.(\ref{variation-trans-action}) actually follows from eq.(\ref{variation-transgression}) can be seen by considering the variation of the Chern-Simons form which is generically given by
$$\delta \mathcal{C}_{2n+1}(A)=\mathfrak{A} _{2n+1}(A,\delta A)+ d\mathfrak{C}_{2n}(A,\delta A)$$
Then on the one hand
\begin{eqnarray}
\delta \mathcal{T}_{2n+1}(A,\overline{A})=\mathfrak{A} _{2n+1}(A,\delta A)-\overline{\mathfrak{A}} _{2n+1}(\overline{A},\delta \overline{A})
+ d[\mathfrak{C}_{2n}(A,\delta A)- \overline{\mathfrak{C}}_{2n}(\overline{A},\delta \overline{A})
+\delta\mathcal{B}_{2n} (  A,\overline{A})]\nonumber
\end{eqnarray}
and on the other hand
$$\delta\int _{\cal M} \mathcal{C}_{2n+1}=
\int _{\cal M}\mathfrak{A} _{2n+1}+ \int _{\partial {\cal M}}\mathfrak{C}_{2n}$$
and therefore 
$$\delta I_{Trans}=\int _{\cal M}\mathfrak{A} _{2n+1}(A,\delta A)-\int _{\overline{\cal M}}\overline{\mathfrak{A}} _{2n+1}(\overline{A},\delta \overline{A})
+ \int _{\partial {\cal M}}[\mathfrak{C}_{2n}(A,\delta A)- \overline{\mathfrak{C}}_{2n}(\overline{A},\delta \overline{A})
+\delta\mathcal{B}_{2n} (  A,\overline{A})]$$ which proves our assertion.}.

A well defined action principle requires a well defined action, with well defined fundamental fields (dynamical variables), and suitable boundary conditions such that the 
variation of the action yields the sum of a bulk term, which vanishes as a result of the field equations, plus a boundary term that vanishes as a result of both the field equations and the boundary conditions.
In the case of the action $I_{Trans}$ at least two natural choices present itself for the dynamical variables:\\
i. Both $A$ and $\overline{A}$ are taken to be dynamical fields.\\
ii. While $A$ is taken to be a dynamical field $\overline{A}$ is considered a fixed non-dynamical background.\\
A third non equivalent choice, leading to gauged Wess-Zumino actions, is to
consider $A$ and $\overline{A}$ as related by a gauge transformation whose 
parameters are independent dynamical variables. We will not discuss this third possibility here.\\
In the case i. we see from eq.(\ref{variation-trans-action}) that the field equations are
\begin{eqnarray}
<F^nG_{\alpha}>=0~~~,~~~<\overline{F}^nG_{\alpha}>=0 \label{field-equations-trangression-i}
\end{eqnarray}
while in the case ii., where $\delta \overline{A}=0$, the field equations are just
\begin{eqnarray}
<F^nG_{\alpha}>=0  \label{field-equations-trangression-ii}
\end{eqnarray}
In both cases it should happen that the boundary conditions are such that, when the field equations hold,
the boundary contribution to the variation of the action $-n(n+1)\int _{\partial {\cal M}}\{\int _0^1dt<\Delta AF_t^{n-1}\delta A_t>\}$ vanishes. We will address these issues in the next section for the particular case of 
Transgression and Chern-Simons AdS gravity.

\section{Chern-Simons and Transgression Gravity}

For the AdS group in dimension $d=2n+1$ the gauge connection is given by\footnote{A gauge
connection has dimensions of $(length)^{-1}$, so it must be  
$A=\frac{\omega ^{ab}}{2}J_{ab}+\frac{e^a}{l}P_a$ where $l$ is the 'AdS radius'.
I set $l=1$ trough all the present paper. It is
easy to reintroduce $l$ using dimensional
analysis, if necessary.}
$A=\frac{\omega ^{ab}}{2}J_{ab}+e^aP_a$ where $\omega ^{ab}$ is the spin connection, $e^a$ is the vielbein and $J_{ab}$ and $P_a$
are the generators of the AdS group (for Lorentz transformations and
translations respectively).  One possible symmetrized trace, and the only one I will consider in this paper, is that which is non zero only for one $P$ generator and $n$  $J$ generators, with values
\begin{equation}
<J_{a_{1}a_{2}}...J_{a_{2n-1}a_{2n}}P_{a_{2n+1}}>=\kappa\frac{2^{n}}
{(n+1)}\epsilon_{a_{1}...a_{2n+1}}
\end{equation}
where $\kappa$ is a constant, which together with the AdS group parameter $l$ ("AdS radius") will characterize the theories.
In addition to the basis of the algebra spanned by the generators $P_a$ and $J_{ab}$ we will
use a basis spanned by the generators $P_1$, $P_i$, $P_i+J_{1i}$ and $P_i-J_{1i}$, 
with $i$ an index taking any allowed value but 1. For this generators the only non 
zero values of the symmetrized trace are 
\begin{eqnarray}
<J_{i_{1}i_{2}}...J_{i_{2n-1}i_{2n}}P_{1}>=\kappa\frac{2^{n}}
{(n+1)}\epsilon_{1i_{1}...i_{2n}}\\
<J_{i_{1}i_{2}}...J_{i_{2n-1}i_{2n-2}}(P_{i_{2n-1}}\pm J_{1i_{2n-1}})
(P_{i_{2n}}\mp J_{1i_{2n}})>=\pm\kappa\frac{2^{n+1}}
{(n+1)}\epsilon_{1i_{1}...i_{2n}}
\end{eqnarray}
Notice in particular that
\begin{eqnarray}
<J_{i_{1}i_{2}}...J_{i_{2n-1}i_{2n-2}}(P_{i_{2n-1}}\pm J_{1i_{2n-1}})
(P_{i_{2n}}\pm J_{1i_{2n}})>= 0
\end{eqnarray}
The transgression for the AdS group  is\footnote{In what follows I will use a
compact notation where $\epsilon$ stands for the Levi-Civita symbol
$\epsilon _{a_1...a_d}$ and wedge products of differential forms are
understood, as it was done in Refs.\cite{motz2,motz1,motz3}. For instance: $\epsilon Re^{d-2}\equiv \epsilon
_{a_1a_2....a_d}R^{a_1a_2}\wedge e^{a_3}\wedge ...\wedge
e^{a_{d-2}}$,  $(\theta ^2)^{ab}=\theta ^a_c\wedge\theta ^{cb}$.} \cite{motz3}
\begin{equation}
{\cal T}_{2n+1}= \kappa \int _0^1dt \epsilon (R+t^2e^2)^ne -\kappa
\int _0^1dt \epsilon (\overline{R}+t^2\overline{e}^2)^n\overline{e}
+d~B _{2n}
\end{equation}
where
\begin{equation}
B _{2n}=-\kappa n\int_0^1dt\int_0^1ds~\epsilon\theta e_t
\left\{t R+(1-t)\overline{R}-t(1-t)\theta ^2+s^2e_t^2 \right\}^{n-1}
\end{equation}
Here $e ^a$ and $\overline{e}^a$ are the two vielbeins and $\omega
^{ab} $ and $\overline{\omega}^{ab}$ the two spin connections,
$R=d\omega +\omega ^2$ and
$\overline{R}=d\overline{\omega}+\overline{\omega}^2$ are the
corresponding curvatures, $\theta =\omega -\overline{\omega}$ and
$e_t=te+(1-t)\overline{e}$. Written in a more compact way
\begin{equation}
B _{2n}=-\kappa n\int_0^1dt\int_0^1ds~\epsilon\theta e_t
\overline{R} _{st}^{n-1}
\end{equation}
where
$$\overline{R} _{st}=t R+(1-t)\overline{R}-t(1-t)\theta ^2+s^2e_t^2$$

The action for transgressions for the AdS group is chosen to be
\cite{motz3}
\begin{equation}
I_{Trans}= \kappa \int _{\cal M}\int _0^1dt \epsilon (R+t^2e^2)^ne
-\kappa\int_{\overline{{\cal M}}} \int _0^1dt \epsilon
(\overline{R}+t^2\overline{e}^2)^n\overline{e} +\int _{\partial {\cal
M} }B _{2n}\label{action-trangresion-ads}
\end{equation}
where ${\cal M}$ and $\overline{\cal M}$ are two manifolds with a
common boundary, that is $\partial {\cal M}\equiv\partial
\overline{{\cal M}}$. Notice that, as said in the previous section, this is a generalization from the
simpler case where ${\cal M}\equiv\overline{\cal M}$.  

We have, as it was said before, the two natural choices of  either regarding both $A$  and $\overline{A}$ as 
dynamical fields, or regarding one of them (lets say $\overline{A}$) as a non independent background, which we will see is given in terms of the boundary data on $A$.

The field equations derived from extremizing the action of eq.(\ref{action-trangresion-ads}) are\footnote{leaving aside for later the question of requiring the vanising of the boundary part of the variation when the field equations and the boundary conditions hold}  
$<F^nG_{\alpha}>=0$ and $<\overline{F}^nG_{\alpha}>=0$, or (see for instance \cite{motz3}) 
\begin{eqnarray}
\epsilon (R+e^2)^n=0~~~,~~~\epsilon (R+e^2)^{n-1}T=0 \\
\epsilon (\overline{R}+\overline{e}^2)^n=0~~~,\label{field-equations-transgression-ads}
~~~\epsilon (\overline{R}+\overline{e}^2)^{n-1}\overline{T}=0
\end{eqnarray}
If $\overline{A}$ is taken to be non dynamical only the first line of the previous equations should hold.

In order to address the questions of the well definiteness of the action principle, and the finiteness  of the conserved charges and the euclidean action we need to be more specific about the asymptotic behaviour of the fundamental fields and the geometry. That is done in the next section.

\section{Space-times with finite AdS gauge curvature at the boundary}
 
\subsection{Asymptotically Locally AdS space-times and Fefferman-Graham metric}

The standard form of the Fefferman-Graham metric for Asymptotically Locally Anti-de Sitter spaces (ALAdS) of dimension $d$ is
\begin{eqnarray}
ds^2=\frac{d\rho ^2}{4\rho ^2}+\frac{1}{\rho}g_{IJ}(x,\rho )dx^Idx^J\label{Fefferman-Graham-metric}
\end{eqnarray}
with $\rho =0$ corresponding to the boundary, where $I,J=0,...,d-1$ but $I,J\neq 1$ (the index 1 corresponds to the "radial" coordinate $\rho $). In the Fefferman-Graham expansion functions $g_{IJ}(x,\rho )$ admit an expansion around $\rho =0$ of the form
\begin{eqnarray}
g_{IJ}(x,\rho ) =\sum _{k=0}^{\infty}g_{IJ}^{(k)}(x,\rho )\rho ^k
\end{eqnarray}
where $g_{IJ}^{(k)}(x,\rho )$ have at most logarithmic divergences and only for $d$ odd (so that the dimension of the boundary is even) and from order $k=\frac{d-1}{2}$. 

If one changes variables so that $dr^2=\frac{d\rho ^2}{4\rho ^2}$ then $dr=\pm d[\frac{1}{2}ln\rho ]$. Choosing the minus sign and the integration constant equal to one we get $r=-\frac{1}{2}ln\rho$ or $\rho =e^{-2r}$, so that when $r\rightarrow +\infty$ then $\rho\rightarrow 0$. In this coordinates, used for instance in Ref.\cite{banados-chandia-ritz}, which we will call "radially simple coordinates", the metric reads
\begin{eqnarray}
ds^2=dr^2+e^{2r}g_{IJ}(x,r)dx^Idx^J\label{FG-Solodukhin}
\end{eqnarray} 
with $g_{IJ}(x,r)\equiv g_{IJ}(x,\rho =e^{-2r} )$.

In what follows I will consider a different situation. While I will study space-times that 
are a perturbation of AdS space-time, 
as in the case of ALAdS space-times admitting a Fefferman-Graham expansion, 
I will look at the weaker condition of just having a finite AdS gauge curvature at the boundary. 
That means that the bulk torsion will in general be non zero, and that we will consider an expansion of the vielbein and the spin connection (assumed at first to be independent), instead of an expansion of the metric.
I will assume the metric has the form of eq.(\ref{FG-Solodukhin}), with a finite $g_{IJ}(x,r )$ 
at the boundary ($r\rightarrow\infty$), but I will derive the information that I will need about 
the form of the expansion from scratch, since I will discuss a more general case.
Expansions of the vielbein and spin connection a la Fefferman-Graham have been considered in refs.\cite{banados-chandia-ritz,banados-miskovic-theisen}, but with a different outlook, as those papers focus on solutions to the field equations and the zero torsion case, rather than the generic off-shell condition of finite asymptotic AdS gauge curvature that we consider below.

\subsection{Vielbein, Spin Connection, Torsion and Curvature}

The vielbein corresponding to this metric are
\begin{eqnarray}
e^1=dr~~~,~~~e^i=e^r\hat{e}^i(x,r)=e^r\hat{e}^i_I(x,r)dx^I
\end{eqnarray}
where $\hat{e}^i_I(x,r)$ satisfy $\eta _{ij}\hat{e}^i_I(x,r)\hat{e}^j_J(x,r)=g_{IJ}(x,r)$ and therefore it is finite at the boundary
$r\rightarrow\infty$.

We will also need the spin connection $\omega ^{ab}$, which we will decompose as $\omega ^{1i}$ and $\omega ^{ij}$. 
The spin connection is not determined by the vielbein, because I will not assume that the torsion is zero.  

In what follows we will consider several p-forms with different structure of indices, which go to a finite limit when $r\rightarrow\infty$.
Those p-forms admit a generic asymptotic expansion at the boundary of the form
\begin{eqnarray}
f^{abc...}(x,r)=f^{abc...}_{\infty }(x)+\sum _{k=1}^{\infty}f_{(k)}^{abc...}(x)\mathcal{P}_{(k)}(r)e^{-kr} 
\end{eqnarray}
where the $f_{(k)}^{abc...}(x)$´s are finite differential forms of the same order than $f^{abc...}(x,r)$ where the $x$ dependence resides,
$\mathcal{P}_{(k)}(r)$ is a function in $r$ which could go to infinity, but only in such a way that $\lim _{r\to\infty}\mathcal{P}_{(k)}(r)e^{-kr}=0$, and also that if $q>k$ then $\mathcal{P}_{(q)}(r)e^{-qr} $ vanishes 
faster than $\mathcal{P}_{(k)}(r)e^{-kr}$ when $r\rightarrow\infty$, independently of the 
$\mathcal{P}_{(k)}(r)$ and $\mathcal{P}_{(q)}(r)$ factors.

That this should be so follows from organizing the dependence of the vanishing part in $r$ as powers of the 
small parameter $e^{-r}$,  but taking in account 
the possibility of factors that are possibly divergent functions of  $r$  at each order. 
 
The detailed form of the functions $\mathcal{P}_{(k)}(r)$ will only be necessary for us for the very first orders of the expansion, for which we will see they are equal to 1. 
Characterizing the generic form of these functions is an interesting problem, which possibly has a very simple solution. A first guess could be that when $r\rightarrow\infty$ then $\mathcal{P}_{(k)}(r)$ could diverge (if it diverges at all) at most as  $r^{s_k}[\ln r]^{p_k}$, where $s_k$ and $p_k$ are positive or zero real numbers. 
However, considering that in the particular case of zero torsion and asymptotically AdS space-times (Fefferman-Graham expansion) those functions are just equal to 1, except (and that only for odd-dimensional space-times) at certain order where it is proportional to $r$ (wich in the $\rho$ variable expansion is a logarithmic factor), it seems that the answer could be quite simpler.
 
We will often distinguish between the part that is finite at the boundary $f^{abc...}_{\infty }(x)$ and the part that vanishes at the boundary
$\sum _{k=1}^{\infty}f_{(k)}^{abc...}(x)\mathcal{P}_{(k)}(r)e^{-kr} $.
We will identify parts of the fields of interest that diverge, those that are finite and those that vanish at the boundary as follows

\begin{eqnarray}
e^i(x,r)=e^r\hat{e}^i(x,r)= e^r[\hat{e}^i_{\infty}(x)+\beta ^i(x,r)]\\
\omega ^{1i}=-e^rk^i(x,r)=-e^r[\hat{e}^i_{\infty}(x)+\alpha ^i(x,r)]\\
\omega ^{ij}(x,r)=\hat{\omega }_{\infty} ^{ij}(x)+\gamma^{ij}(x,r)+\omega ^{ij}_r(x,r)dr
\end{eqnarray}

Several remarks are in order: \\
(i) Because of the form of $\hat{e}^i(x,r)$ then $\beta ^i(x,r)=\beta ^i_Idx^I$ is a 1-form with no component along $dr$ which, 
by definition of $\hat{e}^i_{\infty}(x)$, vanishes at the boundary.\\
(ii) The expression for $\omega ^{1i}$  is given with the benefit of hindsight, as the 1-form 
$\alpha ^i(x,r)$  (with components along $dx ^I$ and $dr$) does not need in principle to vanish at 
the boundary, but we will see that it does.\\
 (iii) The 1-form $\gamma^{ij}=\gamma^{ij}_Idx^I$ is taken to have only $dx^I$ components, 
while $\omega ^{ij}_r$ is the pure number component of $\omega ^{ij}$ along $dr$.\\ 
(iv) We will see that $\omega ^{ij}$ must be finite 
at the boundary, with $\hat{\omega }_{\infty} ^{ij}$ finite and $\gamma^{ij}$ and $\omega ^{ij}_r$ 
vanishing as $r\rightarrow\infty$.  

We will write the exterior derivative distinguishing the radial coordinate from the others as
\begin{eqnarray}
d=\hat{d}+d_r~~~with~~~\hat{d}=dx^I\frac{\partial ~}{\partial x^I}=dx^I\partial _I~~~and~~~d_r=dr\frac{\partial ~}{\partial r}=dr\partial _r
\end{eqnarray}

Remember that if the AdS gauge connection is $A=\frac{1}{2}\omega ^{ab}J_{ab}+e^aP_a$  then the AdS 
gauge curvature is $F=dA+A^2=\frac{1}{2}[R^{ab}+e^ae^b]J_{ab}+T^aP_a$, where the curvature $R^{ab}$ 
and the torsion $T^a$ are defined by $R^{ab}=d\omega ^{ab}+\omega ^a_{~c}\omega^{cb}$ and 
$T^a=de^a+\omega ^{a}_{~b}e^b$ respectively. Therefore $F^{ab}=R^{ab}+e^ae^b$ and $F^a=T^a$.
 
For the components $F^a$ of the AdS curvature along the $P_a$ generators, which are the components of the torsion,  we get
\begin{eqnarray}
F^1=T^1=-e^{2r}[\zeta _i\hat{e}^i_{\infty}+\frac{1}{2}\zeta _i\tau ^i]\label{F1}\\
F^i=T^i=e^r[\hat{T}^i_{\infty}+\hat{D}_{\infty}\beta ^i+\gamma ^i_{~j}\beta ^j+\gamma ^i_{~j}\hat{e}^j_{\infty}+dr(D_r\beta ^i-\zeta ^i+\omega ^i _{r~j}\hat{e}^i_{\infty} )]\label{Fi}
\end{eqnarray}
where $\zeta ^i\equiv \alpha ^i-\beta ^i$ and $\tau ^i\equiv \alpha ^i+\beta ^i$, 
$\hat{T}^i_{\infty}\equiv\hat{d}\hat{e}^i_{\infty}+\hat{\omega}^i_{\infty ~j}\hat{e}^j_{\infty}\equiv \hat{D}_{\infty}\hat{e}^i_{\infty}$ is the boundary torsion, and $\hat{D}_{\infty}\beta ^i\equiv\hat{d}\beta ^i+\hat{\omega}^i_{\infty ~j}\beta ^j $ and 
$D_r\beta ^i\equiv\partial _r\beta ^i+\omega ^i _{r~j}\beta ^j$ are a sort of boundary and radial 
covariant derivatives respectively.

For the components $F^{ab}$ of the AdS gauge curvature we get
\begin{eqnarray}
F^{ij}=\hat{R}^{ij}_{\infty}+ \hat{D}_{\infty}\gamma ^{ij}+
\gamma ^i_{~k}\gamma ^{kj} -e^{2r}[\zeta ^i\hat{e}^j_{\infty}+
\hat{e}^i_{\infty}\zeta ^j+\frac{1}{2}(\zeta ^i\tau ^j+\tau ^i\zeta ^j)]+\nonumber\\
+dr[D_r\gamma ^{ij}-\hat{D}_{\infty}\omega _r^{ij}]\label{Fij}  \\
F^{1i}= -e^r[\hat{T}^i_{\infty}+\hat{D}_{\infty}\alpha ^i+\gamma ^i_{~j}\alpha ^j+\gamma ^i_{~j}\hat{e}^j_{\infty}+
dr(D_r\alpha ^i+\zeta ^i+\omega ^i _{r~j}\hat{e}^i_{\infty} )]\label{F1i}
\end{eqnarray}
where $\hat{D}_{\infty}\gamma ^{ij}=\hat{d}\gamma ^{ij}+\hat{\omega }^{i} _{\infty~ k}\gamma ^{kj}+\hat{\omega }^{j} _{\infty~k}\gamma ^{ik}$, 
$\hat{D}_{\infty}\omega _r^{ij}=\hat{d}\omega _r^{ij}+\hat{\omega }^{i} _{\infty~ k}\omega _r^{kj}+\hat{\omega }^{j} _{\infty~k}\omega _r^{ik}$ and 
$D_r\gamma ^{ij}=\partial _r\gamma ^{ij}+\omega _{r~k}^{i}\gamma ^{kj}+\omega _{r~k}^{j}\gamma ^{ik}$. Some care is required handling the minus signs that appear as a result of pulling the $dr$ to the left, in some cases trough 1-forms.
 
It is actually more natural to write everything in terms of $\zeta ^i$ and $\tau ^i$ rather than $\alpha ^i$ and $\beta ^i$, because the former 
fields vanish at a different rate towards the boundary, as we will show in the next subsection. This goes hand in hand with 
changing the basis of the AdS gauge algebra replacing $P_i$ and $J_{1i}$ by $P_i+J_{1i}$ and $P_i-J_{1i}$. We get for $F$
\begin{eqnarray}
F=\frac{1}{2}F^{ij}J_{ij}+F^1P_1+\nonumber\\
e^r[\hat{T}^i_{\infty}+\frac{1}{2}(\hat{D}_{\infty}\tau ^i+\gamma ^i_{~j}\tau ^j)+ 
\gamma ^i_{~j}\hat{e}^j_{\infty} +dr(\frac{1}{2}D_r\tau ^i+\omega ^i _{r~j}\hat{e}^i_{\infty})](P_i-J_{1i})-\nonumber\\
-e^r[\frac{1}{2}(\hat{D}_{\infty}\zeta ^i+\gamma ^i_{~j}\zeta ^j) 
 +dr(\frac{1}{2}D_r\zeta ^i+\zeta ^i)](P_i+J_{1i})\label{F-total}
\end{eqnarray}
where $F^{ij}$ is given by eq.(\ref{Fij}) and $F^1$ is given by eq.(\ref{F1}).

\subsection{Conditions on the fields for a finite AdS gauge curvature at the boundary}

The main idea of this section is to find the conditions imposed on the fields (vielbein and spin connection)
by requiring the AdS gauge curvature to be finite at the boundary $r\rightarrow\infty$. It is important to remember that te components of the vielbein and the spin connection are independent, therefore we cannot generically assume cancellations between terms involving different fields.\\ 
Requiring $F^{ab}$ to be finite and using the equations given in the previous subsection we obtain the following conclusions:\\
(i) From eq.(\ref{Fij}) it results that $\zeta ^i(x,r)=e^{-2r}\hat{\zeta }^i(x,r)$ with $\hat{\zeta }^i(x,r)$ finite at the boundary.\\
(ii) From eq.(\ref{Fi}) or eq.(\ref{F1i}) we see that the torsion of the boundary itself must vanish $\hat{T}^i_{\infty}(x)=0$,
which in turn implies that the boundary spin connection is not independent but it is instead a functional of the boundary vielbein 
$\hat{\omega }^{ij} _{\infty}(x)=\hat{\omega }^{ij} _{\infty}[\hat{e}^i_{\infty}(x)]$. This 
implies $\hat{\omega }^{ij} _{\infty}$ is finite if $\hat{e}^i_{\infty}$ is finite, as assumed.\\
(iii) If we consider the coefficient of $P_i-J_{1i}$ in eq.(\ref{F-total}) we see that $\tau ^i(x,r)=e^{-r}\hat{\tau}^i(x,t)$, where 
$\hat{\tau}^i$ is finite at the boundary.\\
(iv) Taking in account the previous points, and considering the coefficient of $P_i-J_{1i}$ we see that
$\gamma ^{ij}(x,r)=e^{-r}\hat{\gamma } ^{ij}(x,r)$ with $\hat{\gamma } ^{ij}$ finite at the boundary, and that
$\omega _r ^{ij}(x,r)=e^{-r}\hat{\omega }_r ^{ij}(x,r)$ with $\hat{\omega}_r^{ij}$ finite at the boundary.\\ 
The gauge potential that results from requiring a finite gauge curvature is
\begin{eqnarray}
A=e^r\hat{e}^i_{\infty}(P_i-J_{1i})+\nonumber\\
+\frac{1}{2}\hat{\omega}_{\infty}^{ij}J_{ij}+\frac{1}{2}\hat{\tau}^i(P_i-J_{1i})+dr~P_1+\nonumber\\
+\frac{1}{2}e ^{-r}[\hat{\gamma}^{ij}+dr\hat{\omega}_r^{ij}]J_{ij}-\frac{1}{2}e ^{-r}\hat{\zeta}^i(P_i+J_{1i})\label{asymptotic-A}
\end{eqnarray}
where, at the boundary, the first line diverges as $e^r$, the second line is finite and the third line vanishes as $e^{-r}$.

For the gauge curvature we have
\begin{eqnarray}
F=\frac{1}{2}F^{ij}J_{ij}+F^1P_1+\nonumber\\
+\{ \frac{1}{2}\hat{D}_{\infty}\hat{\tau } ^i+ 
\hat{\gamma } ^i_{~j}\hat{e}^j_{\infty}+dr[\frac{1}{2}(\partial _r\hat{\tau } ^i- \hat{\tau } ^i)+\hat{\omega } ^i_{r~j}\hat{e}^j_{\infty} ]\}(P_i-J_{1i})+\nonumber\\
 + \frac{1}{2}e^{-r}\{ \hat{\gamma }^i_{~j}\hat{\tau } ^j+dr \hat{\omega } _{rj}^i\hat{\tau}^j\} (P_i-J_{1i})
-\frac{1}{2}e^{-r}[\hat{D}_{\infty}\hat{\zeta } ^i
 +dr \partial _r\hat{\zeta }^i]  (P_i+J_{1i})-\nonumber\\
-\frac{1}{2} e^{-2r}\{ \hat{\gamma } ^i_{~j}\hat{\zeta } ^j+dr\hat{\omega } _{rj}^i\hat{\zeta }^j  \}(P_i+J_{1i}) \label{asymptotic-F}
\end{eqnarray}
where, at the boundary, the second line is finite, the third line goes to zero as $e^{-r}$ and the fourth 
line vanishes as $e^{-2r}$ . The first line of the previous equation contains terms of several orders at the boundary (finite and vanishing), which are
explicitly given by
\begin{eqnarray}
F^{ij}=\hat{R}^{ij}_{\infty}-  
\hat{\zeta }^i\hat{e}^j_{\infty}-\hat{e}^i_{\infty}\hat{\zeta }^j+\nonumber\\
+e^{-r}[\hat{D}_{\infty}\hat{\gamma } ^{ij}-\frac{1}{2}(\hat{\zeta }^i\hat{\tau }^j+\hat{\tau }^i\hat{\zeta }^j)
+dr(\partial _r\hat{\gamma }^{ij}-\hat{D}_{\infty}\hat{\omega }_r^{ij})]\nonumber\\
+e^{-2r}\{ \hat{\gamma }^i_{~k}\hat{\gamma }^{kj}+ 
dr[\hat{\omega}_{rk}^i\hat{\gamma}^{kj}+\hat{\omega}_{rk}^j\hat{\gamma}^{ik} ]\} \label{asymptotic-Fij}  \\
F^1=-\hat{\zeta } _i\hat{e}^i_{\infty}-\frac{1}{2}e^{-r}\hat{\zeta }_i\hat{\tau }^i \label{asymptotic-F1} 
\end{eqnarray}
The finite part of $F$ at the boundary is given by
\begin{eqnarray}
F^{(finite)}=\frac{1}{2} [\hat{R}^{ij}_{\infty}-  
\hat{\zeta }^i\hat{e}^j_{\infty}-\hat{e}^i_{\infty}\hat{\zeta }^j] J_{ij}
-\hat{\zeta } _i\hat{e}^i_{\infty} P_1+\nonumber\\
+\{ \frac{1}{2}\hat{D}_{\infty}\hat{\tau } ^i+ 
\hat{\gamma } ^i_{~j}\hat{e}^j_{\infty}+dr[\frac{1}{2}(\partial _r\hat{\tau } ^i- \hat{\tau } ^i)
+\hat{\omega } ^i_{r~j}\hat{e}^j_{\infty} ]\}(P_i-J_{1i})\label{finite-F}
\end{eqnarray}
Notice that $F^{(finite)}$ has no components along $(P_i+J_{1i})$.

\subsection{Particular case of zero AdS gauge curvature at the boundary}

In this subsection we will look at the special cases for which 
$$\lim _{r\to\infty}F^{(finite)}(x,r)=0$$
If the non vanishing part of $\hat{\zeta }^i(x,r)$ is 
$$\hat{\zeta }^i_{\infty}(x)\equiv\lim _{r\to\infty}\hat{\zeta }^i (x,r)$$
is given in the $\hat{e}^i_{\infty}$ basis as $\hat{\zeta }^i_{\infty}(x)=C^i_{~j}(x)\hat{e}^j_{\infty}(x)$ (assuming that it has not component along $dr$, which in general may exist) and the
2-form $\hat{R}^{ij}_{\infty}$ is given in the same basis by 
$\hat{R}^{ij}_{\infty}(x)=\frac{1}{2}\mathcal{R}^{ij}_{kl}(x)\hat{e}^k_{\infty}(x)\hat{e}^l_{\infty}(x)$, 
then the vanishing of the the component of $F^{(finite)}$ along $P_1$ implies that $C_{ij}=C_{ji}$, while the
vanishing of the $J_{ij}$ component implies
\begin{eqnarray}
\frac{1}{2}\mathcal{R}^{ij}_{kl}-C^i_{~[k}\delta ^j_{l]}-C^j_{~[l}\delta ^i_{k]}=0
\end{eqnarray}
where $C^i_{~[k}\delta ^j_{l]}=\frac{1}{2}[C^i_{~k}\delta ^j_{l}-C^i_{~l}\delta ^j_{k}]$. This equation is solved, for a boundary dimension $D>2$, by contracting
upper and lower indices, yielding
\begin{eqnarray}
C^i_{~k}=\frac{1}{(D-2)}[\mathcal{R}^{i}_{k}-\frac{\mathcal{R}}{2(D-1)}\delta ^{i}_{k}]\label{C-zeta}
\end{eqnarray}
where the boundary Ricci tensor $\mathcal{R}^{i}_{k}$ is the result of contracting $j$ and $l$ in the boundary Riemann tensor
$\mathcal{R}^{ij}_{kl}$, while the Riemann curvature scalar is $\mathcal{R}=\mathcal{R}^{i}_{i}$. The tensor $C_{ij}$ is symmetric, as required, and it is the well known {\it Schouten tensor}. We 
conclude that $\hat{\zeta }^i_{\infty}(x)$ is fully determined and explicitly calculable in terms of $\hat{e}^i_{\infty}(x)$
if we require $F$ to vanish at the boundary \footnote{ In the case of $D=2$ the previous expression for $C^i_{~k}$ is evidently not valid. The index $i,~ j,~ k,~ l$ can only take the values 0 and 2, and one can verify that the only condition required to make $\frac{1}{2}\mathcal{R}^{ij}_{kl}-C^i_{~[k}\delta ^j_{l]}-C^j_{~[l}\delta ^i_{k]}=0$ is that
$ C=C^0_{~0}+C^2_{~2}=\frac{\mathcal{R}}{2}$. In this case $C_{02}$ and $C_{20}$ are not determined, but
the vanishing of the torsion requires $C_{02}=C_{20}$. The condition on 
$\hat{\tau}^i$, $\hat{\gamma}^{ij}$ and $\hat{\omega}_r^{ij}$ is the same as for any $D$, 
and can be satisfied, as a sufficient but not necessary condition, by setting these fields to zero at the boundary.}.

The vanishing of the remaining piece of  $F^{(finite)}$, which lies along $(P_i-J_{1i})$, requires
\begin{eqnarray}
\frac{1}{2}\hat{D}_{\infty}\hat{\tau } ^i+ 
\hat{\gamma } ^i_{~j}\hat{e}^j_{\infty}+dr[\frac{1}{2}(\partial _r\hat{\tau } ^i- \hat{\tau } ^i)
+\hat{\omega } ^i_{r~j}\hat{e}^j_{\infty} ]=0 \label{tau-cero-F}
\end{eqnarray}
These equations do not fully determine the boundary values of $\hat{\tau } ^i$, 
$\hat{\gamma } ^i_{~j}$ and $\hat{\omega } ^i_{r~j} $, but give relationships between these fields and
the radial derivative of $\hat{\tau } ^i$. An immediate solution of eq.(\ref{tau-cero-F}), which we will use below, is
$\hat{\tau } ^i=0$, $\hat{\gamma } ^i_{~j}=0$ and $\hat{\omega } ^i_{r~j} =0$.
 
Of course the most special case, for which $F=0$ everywhere, is the AdS space-time, 
which in several of its presentations is naturally describable within the previous framework.
 
\section{Action Principle and Boundary Conditions I: Backgrounds}

\subsection{Goal of this section and the next}

The action functional is defined as the integral of the Lagrangian density (or rather d-form)
in a space-time region bounded by two "constant time" space-like hyper-surfaces $\Sigma _{t_0}$
and $\Sigma _{t_1}$ and the spacial infinite ($r\rightarrow\infty$) boundary 
between those two times $\Sigma _{\infty}$. 
The general variation of the action consists of a bulk and a boundary contribution. 
The condition for the vanishing of the bulk contribution are just the field equations of the 
theory. It is said say that one has a well defined action principle if, for field configurations
satisfying the field equations, the boundary contribution also vanishes when 
suitable boundary conditions are imposed. In that case the action is a true extremum when 
the field equations and boundary conditions hold, which allows us to consider 
that configuration as a true classical limit
of the quantum theory, in a path integral formulation, when the conditions 
for taking the saddle point
 approximation hold\footnote{Essentially that the action should be much larger than $\hbar$.}.

We will assume that both the vielbein 
and the spin connection are kept fixed at $\Sigma _{t_0}$ and $\Sigma _{t_1}$, 
and therefore that $\delta A=0$ at those manifolds. From eq.[\ref{variation-transgression}] we see that
the boundary contribution from those manifolds vanishes. Alternatively, we will consider also the euclidean case with 
periodic euclidean time, in which case $\Sigma _{t_0}$
and $\Sigma _{t_1}$ are identified and the only boundary is $\Sigma _{\infty}$. In both cases 
we just need to ensure the vanishing of the boundary contribution to the variation 
of the action at $\Sigma _{\infty}$. We will not assume the whole vielbein and spin connection are given at
$\Sigma _{\infty}$, but only its intrinsic vielbein and the asymptotic behavior discussed 
in the previous section will be assumed.

At the quantum level, which we will only consider in a heuristic way, we have amplitudes
between given configurations for the vielbein and spin connection (modulo gauge) at $\Sigma _{t_0}$
and $\Sigma _{t_1}$ given by a path integral summing contributions of all configurations 
(modulo gauge) interpolating between
those and satisfying the boundary and asymptotic conditions at $\Sigma _{\infty}$ weighted with the standard 
Feynman factor\footnote{For generic gravity theories in a second order formulation 
suitable boundary terms are required in order
to guarantee the additivity of the action, that in turn is required to
 ensure the right composition properties 
of quantum amplitudes (see \cite{myers-1,muller-hoissen-1,muller-hoissen-2,myers-2}). 
That problem results from the fact that the configuration at a given time is specified by giving the metric at that time, 
but the normal derivative of the metric is not determined. 
In our case both the vielbein and the (independent) spin connection are given in specifying the 
configuration at any given time, and the action is therefore additive as it is.}. 

Notice that for arbitrary field 
configurations at $t_0$ and $t_1$ there is not in general a solution of the field equations
with the given boundary conditions that interpolates between them\footnote{The problem of identifying the degrees of 
freedom and general conditions for the the partial differential equations [\ref{field-equations-transgression-ads}] to be solvable (the Cauchy problem for those PDEs) is remarkably subtle and complex, and has been 
studied in refs.\cite{banados-garay-henneaux-1,banados-garay-henneaux-2,miskovic-troncoso-zanelli,miskovic-zanelli}. }. 

We will also show that the conserved charges and euclidean action (and hence thermodynamic quantities) 
computed with the action principles  and asymptotic behavior considered are finite without 
the need for further regularization.

\subsection{Basic ingredients and setup}
We have two configurations $A$ and $\overline{A}$, each with its own vielbein and spin connection (not over lined and over lined respectively), both satisfying the asymptotic condition of having a finite AdS gauge curvature,  characterized therefore  by $\hat{e}^i_{\infty}$, $\hat{\omega}^{ij}_{\infty}$,
$\hat{\zeta}^i$, $\hat{\tau}^i$, $\hat{\gamma}^{ij}$, $\hat{\omega}_r^{ij}$ and $\hat{\overline{e}}^i_{\infty}$, $\hat{\overline{\omega}}^{ij}_{\infty}$,
$\hat{\overline{\zeta}}^i$, $\hat{\overline{\tau}}^i$, $\hat{\overline{\gamma}}^{ij}$, $\hat{\overline{\omega}}_r^{ij}$.
We will furthermore require that the boundary vielbein is the same for both configurations, that is 
$\hat{e}^i_{\infty}=\hat{\overline{e}}^i_{\infty}$, 
which in turn implies $\hat{\omega}^{ij}_{\infty}=\hat{\overline{\omega}}^{ij}_{\infty}$. 

In what follows the subindex $t$ implies interpolation between the non overlined and 
the overlined magnitudes with parameter $t$, $0\leq t\leq 1$. For instance $A_t=tA+(1-t)\overline{A}$, 
as before. 

We will need
\begin{eqnarray}
A_t=e^r\hat{e}^i_{\infty}(P_i-J_{1i})+\nonumber\\
+\frac{1}{2}\hat{\omega}_{\infty }^{ij}J_{ij}+\frac{1}{2}\hat{\tau}_t^i(P_i-J_{1i})+dr~P_1+\nonumber\\
+\frac{1}{2}e ^{-r}[\hat{\gamma}_t^{ij}+dr\hat{\omega}_{r~t}^{ij}]J_{ij}-\frac{1}{2}e ^{-r}\hat{\zeta}_t^i(P_i+J_{1i})\label{asymptotic-A_t}
\end{eqnarray}

\begin{eqnarray}
\delta A_t= e^r\delta\hat{e}^i_{\infty} (P_i-J_{1i})+\nonumber\\
+\frac{1}{2}\delta\hat{\omega}_{\infty}^{ij} J_{ij}+\frac{1}{2}\delta\hat{\tau}_t^i(P_i-J_{1i})+\nonumber\\
+\frac{1}{2}e ^{-r}[\delta\hat{\gamma}_t^{ij}+dr\delta\hat{\omega}_{r~t}^{ij}]J_{ij}
-\frac{1}{2}e ^{-r}\delta\hat{\zeta}_t^i(P_i+J_{1i})\label{delta-A_t}
\end{eqnarray}
It is important to notice that in the previous expression $\delta\hat{\omega}_{\infty}^{ij}$ is not independent of $\delta\hat{e}^i_{\infty}$, as the condition $\hat{T}^i_{\infty}=0$ implies that the boundary vielbein determines the boundary spin connection. Furthermore, we will impose Dirichlet boundary conditions on the vielbein, meaning that the field equations result from requiring that the action is extremal for variations that keep $\delta\hat{e}^i_{\infty}=0$, which in turn imply $\delta\hat{\omega}_{\infty}^{ij}=0$ for those variations.

\begin{eqnarray}
F_t=\frac{1}{2}F_t^{ij}J_{ij}+F_t^1P_1+\nonumber\\
+\{ \frac{1}{2}\hat{D}_{\infty}\hat{\tau }_t ^i+ 
\hat{\gamma } ^i_{t~j}\hat{e}^j_{\infty}+dr[\frac{1}{2}(\partial _r\hat{\tau }_t ^i- 
\hat{\tau }_t ^i)+\hat{\omega } ^i_{rt~j}\hat{e}^j_{\infty} ]\}(P_i-J_{1i})+\nonumber\\
 + \frac{1}{2}e^{-r}\{ \hat{\gamma }^i_{t~j}\hat{\tau }_t ^j+dr \hat{\omega } _{rt~j}^i\hat{\tau}_t^j\} (P_i-J_{1i})
-\frac{1}{2}e^{-r}[\hat{D}_{\infty}\hat{\zeta }_t ^i
 +dr \partial _r\hat{\zeta }_t^i]  (P_i+J_{1i})-\nonumber\\
-\frac{1}{2} e^{-2r}\{ \hat{\gamma } ^i_{t~j}\hat{\zeta }_t ^j+dr\hat{\omega } _{rt~j}^i\hat{\zeta }_t^j  \}(P_i+J_{1i}) \label{F_t-backgrounds}
\end{eqnarray}

\begin{eqnarray}
F_t^{ij}=\hat{R}^{ij}_{\infty}-  
\hat{\zeta }_t^i\hat{e}^j_{\infty}-\hat{e}^i_{\infty}\hat{\zeta }_t^j+\nonumber\\
+e^{-r}[\hat{D}_{\infty}\hat{\gamma }_t ^{ij}-\frac{1}{2}(\hat{\zeta }_t^i\hat{\tau }_t^j+\hat{\tau }_t^i\hat{\zeta }_t^j)
+dr(\partial _r\hat{\gamma }_t^{ij}-\hat{D}_{\infty}\hat{\omega }_{t~r}^{ij})]\nonumber\\
+e^{-2r}\{ \hat{\gamma }^i_{t~k}\hat{\gamma }_t^{kj}+ 
dr[\hat{\omega}_{rt~k}^i\hat{\gamma}_t^{kj}+\hat{\omega}_{rt~k}^j\hat{\gamma}_t^{ik} ]\} \label{Fij_t-backgrounds}  \\
F_t^1=-\hat{\zeta } _{t~i}\hat{e}^i_{\infty}-\frac{1}{2}e^{-r}\hat{\zeta }_{t~i}\hat{\tau }_t^i \label{F1_t-backgrounds} 
\end{eqnarray}

\begin{eqnarray} 
\Delta A =\frac{1}{2}\Delta \hat{\tau}^i(P_i-J_{1i})\nonumber\\
+\frac{1}{2}e ^{-r}[\Delta\hat{\gamma}^{ij}+dr\Delta\hat{\omega}_r^{ij}]J_{ij}-\frac{1}{2}e ^{-r}\Delta\hat{\zeta}^i(P_i+J_{1i})\label{Delta-A-backgrounds}
\end{eqnarray}

\subsection{Action Principle}

The goal of this section is to prove that if:\\
i. the Field Equations of eq.(\ref{field-equations-trangression-i}) hold,\\ 
ii. the field configurations considered are those that yield a finite AdS 
gauge curvature asymptotically, given by eq.(\ref{asymptotic-A}) and \\
iii. The vielbein satisfies Dirichlet boundary conditions $\delta\hat{e}^i_{\infty}=0$,\\
then the action is an extremum, meaning that its variation subjet to those conditions is zero.

The previous statement is equivalent to saying that the AdS Transgression eq.(\ref{action-trangresion-ads}) action for those field configurations and boundary conditions provides a well defined action principle and a well posed variational problem.

To prove that we need the variation of the Transgression eq.(\ref{variation-transgression}). The bulk terms are zero as a result of the Field Equations eq.(\ref{field-equations-trangression-i}), as usual. We are left with the task of proving that the boundary term 
$$-n(n+1)\int _0^1dt<\Delta AF_t^{n-1}\delta A_t>$$
vanishes. To do so we recall that for the symmetrized trace $<...>$ to be non zero it must either:\\
a.  include the generator $P_1$ and as many $J_{ij}$ generators as required with all different indices or\\
b. include both the generators $(P_i-J_{1i})$ and $(P_i+J_{1i})$, and as many $J_{jk}$ generators as required with all different indices (also different from $i$).

With the boundary condition imposed we get
\begin{eqnarray}
\delta A_t=\frac{1}{2}\delta\hat{\tau}_t^i(P_i-J_{1i})+\nonumber\\
+\frac{1}{2}e ^{-r}[\delta\hat{\gamma}_t^{ij}+dr\delta\hat{\omega}_{r~t}^{ij}]J_{ij}
-\frac{1}{2}e ^{-r}\delta\hat{\zeta}_t^i(P_i+J_{1i})\label{delta-A_t-dirichlet}
\end{eqnarray}
We see that both $\delta A_t$ and $\Delta A$ have a finite part only along $(P_i-J_{1i})$, while $F_t$
has at most finite components, in principle, along every generator. Neither of the factors has divergences for $r\rightarrow\infty$. Now, the finite component of $\delta A_t$ along $(P_i-J_{1i})$ must come with a $(P_i+J_{1i})$ either on $\Delta A$ or on one $F_t$ factor. 
In the first case, the component of $\Delta A$
along $(P_i+J_{1i})$ goes as $e^{-r}$ at the boundary, while $F_t$ is finite, therefore that contribution is zero at the boundary. 
In the second case both the component of $F_t$ along $(P_i+J_{1i})$ and the component of $\Delta A$ along $J_{jk}$ go as $e^{-r}$, therefore these contributions are also zero. 
Finally, only $F_t$ has a possible finite component along $P_1$, 
which must come together with $J_{jk}$ components for the rest of the $F_t$'s, 
$\Delta A$ and $\delta A_t$, but for the last two those components go as $e^{-r}$, while for the first they are at most finite, therefore the whole thing vanishes at the boundary. These considerations prove the claim made at the beginning of the subsection.

It is important to emphasize the non trivial character of the previous result, which could be appreciated considering for instance the pure Chern-Simons case. In that case $\overline{A}=0$, and the reader 
can verify that both $\delta A_t$ and $\Delta A$ have components that diverge as $e^{r}$, resulting on a non vanishing boundary contribution to the variation of the action (which does not therefore provide a well defined action principle by itself).

\subsubsection{Two ways: dynamical or non dynamical backgrounds}

The previous analysis applies either if:\\ 
i. we regard both $A$ and $\overline{A}$ as dynamical fields, 
subjected to the condition of having the same boundary data $\hat{e}^i_{\infty}$, or\\
ii. we consider $A$ as fully dynamical but $\overline{A}$ as a non dynamical background.

In the second case the simplest choice corresponds to choosing $\overline{A}$ in such a 
way that its gauge curvature $\overline{F}$ vanishes at the boundary, as discussed at the end of the previous Section. That configuration, which we may call the "AdS vacuum", is the one with the same boundary vielbein and boundary spin connection as $A$,  $\hat{\overline{\zeta }}^i(x)=\overline{C}^i_{~j}(x)\hat{e}^j_{\infty}(x)$
with
\begin{eqnarray}
\overline{C}^i_{~k}=\frac{1}{(D-2)}[\mathcal{R}^{i}_{k}-\frac{\mathcal{R}}{2(D-1)}\delta ^{i}_{k}]
\end{eqnarray}
and 
$\hat{\overline{\tau }} ^i=0$, $\hat{\overline{\gamma }} ^i_{~j}=0$ and $\hat{\overline{\omega }} ^i_{r~j} =0$. Looking at eq.(\ref{asymptotic-F}), eq.(\ref{asymptotic-Fij}) and eq.(\ref{asymptotic-F1}) 
we see that in fact this choice of $\overline{A}$, with $\hat{\overline{\zeta }} ^i(x)$ independent
of $r$, actually yields a $\overline{F}$ that would vanish everywhere, except for a term containing
$e^{-r}\hat{D}_{\infty}\hat{\overline{\zeta }} ^i$ that vanishes at the boundary but not
necessarily everywhere. Yet in some particular cases 
$\hat{D}_{\infty}\hat{\overline{\zeta }} ^i=0$, for instance if the boundary space is of 
constant curvature, because in that case $\hat{\overline{\zeta }} ^i$ is proportional to the
 boundary vielbein with a constant factor, and the boundary torsion is zero. 
Thankfully (as it will be necessary below), and perhaps surprisingly, even if 
$\hat{D}_{\infty}\hat{\overline{\zeta }} ^i$ is not zero this $\overline{A}$ configuration 
is a solution of the field equations. The reason is that in that case 
$\overline{F}=-\frac{1}{2}e^{-r}\hat{D}_{\infty}\hat{\overline{\zeta }} ^i(P_i+J_{1i})$ then, as far as $n>1$, the field equations $<\overline{F}^nG_{\alpha}>=0$
 are satisfied, as it would be required that $\overline{F}$ has components along generators other than $(P_i+J_{1i})$ to make the trace non vanishing\footnote{ As before, the case of $D=2$ (or equivalently $n=1$) is different. Again the previous expression for $\overline{C}^i_{~k}$ is not correct in that case, and the right conditions are
$ \overline{C}=\overline{C}^0_{~0}+\overline{C}^2_{~2}=\frac{\mathcal{R}}{2}$ and  $\overline{C}_{02}=\overline{C}_{20}$. We set
$\hat{\overline{\tau}}^i=\hat{\overline{\gamma}}^{ij}=\hat{\overline{\omega}}_r^{ij}=0$, which takes care of the other condition.\\ However,
an additional complication is that
in the $D=2$ case, if we want that the AdS vacuum satisfies the field equations everywhere we must require $\overline{F}=0$. 
That implies that we must require $\hat{D}_{\infty}\hat{\overline{\zeta }} ^i=0$, yet one can check that, in addition to the previous conditions, that implies $\hat{d}\mathcal{R}=0$, which means that the boundary has constant curvature. It follows that the AdS vacuum is adequate in $D=2$ only if the boundary manifold is of constant curvature. In that case $\hat{\overline{\zeta }} ^i(x)=\frac{\mathcal{R}}{4}\hat{e}^i_{\infty}(x)$.}.

\subsection{Finite Conserved Charges}

Transgression field theories are invariant under both local coordinate transformations 
(or diffeomorphisms) and local gauge transformations. There are conserved Noether currents
associated to both kind of transformations, and in concrete examples both kinds of 
currents turn out to be equivalent. In the present article I will 
only consider charges associated with local coordinate transformations. 
The detailed structure of gauge transformations that preserve the 
required asymptotic form of the fields, gauge charges and holographic anomalies 
will be investigated in a separate work.

The variation of the gauge potentials under diffeomorphisms generated by an infinitesimal vector field $\xi ^{\mu}$ is
\begin{eqnarray}
\delta _{\xi}A  =-{\cal L } _{\xi} A = -D  [ I _{\xi }A ] -I
_{\xi}F= -[I _{\xi}d+d I _{\xi}]A   
\end{eqnarray}
where the contraction operator $I_{\xi }$ is defined by
acting on a p-form $\alpha _p$ as
$$
I_{\xi }\alpha _{p}=\frac{1}{(p-1)!}\xi ^{\nu }\alpha _{\nu \mu
_{1}...\mu _{p-1}}dx^{\mu _{1}}...dx^{\mu _{p-1}}
$$
and being and anti-derivative in the sense that acting on the wedge
product of differential forms $\alpha _{p}$ and $\beta _{q}$ of
order $p$ and $q$ respectively gives $I_{\xi }(\alpha _{p}\beta
_{q})=I_{\xi }\alpha _{p}\beta _{q}+(-1)^{p}\alpha _{p}I_{\xi }\beta
_{q}$.

Noether's Theorem yields (see for instance \cite{motz3}) the conserved current
\begin{equation}
\ast j = ~d Q _{\xi}
\end{equation}
where the conserved charge density is
\begin{eqnarray}
Q _{\xi} = +n(n+1)\int_0^1dt <\Delta A F_t^{n-1} I _{\xi }A _t>\label{Noether-charge-diffeo}
\end{eqnarray}

In order to discuss the finiteness  of such charge density and give explicit expressions for it we need, in addition to the $\Delta A$ and $F_t$ given above,
\begin{eqnarray}
 I _{\xi }  A_t=e^r I _{\xi }  \hat{e}^i_{\infty}(P_i-J_{1i})+\nonumber\\
+\frac{1}{2}I _{\xi }\hat{\omega}_{\infty }^{ij}J_{ij}+\frac{1}{2}I _{\xi }\hat{\tau}_t^i(P_i-J_{1i})+  I _{\xi }( dr)~P_1+\nonumber\\
+\frac{1}{2}e ^{-r}[I _{\xi }\hat{\gamma}_t^{ij}+I _{\xi }  (dr)\hat{\omega}_{r~t}^{ij}]J_{ij}-\frac{1}{2}e ^{-r}I _{\xi }\hat{\zeta}_t^i(P_i+J_{1i})
\end{eqnarray}
Notice that $I _{\xi }( dr)$ is zero unless $\xi=\frac{\partial ~}{\partial r}$ in which case
it is equal to 1.

We will see first that the charge density is finite, by considering the combinations of factors that have non zero symmetrized trace and their behaviour at the boundary. I will work out explicitly just a few cases: \\
i. If $\Delta A$ is along $P_i-J_{1i}$ (finite), $I _{\xi }A _t$ goes along 
$P_i+J_{1i}$ (goes as $e^{-r}$) then the $F_t$'s have to be along $J_{ij}$ (finite), then the whole thing vanishes at the boundary as $e^{-r}$.\\
ii. If $\Delta A$ is along $P_i-J_{1i}$ (finite), $I _{\xi }A _t$ goes along 
$J_{ij}$ (finite) then $n-2$ of the $F_t$'s have to be along $J_{ij}$ (finite) and one of the 
$F_t$'s have to be along $P_i+J_{1i}$ (goes as $e^{-r}$) 
then the whole thing vanishes at the boundary as $e^{-r}$.\\
iii. If one $F_t$ factor is along $P_i-J_{1i}$ (finite), one $F_t$ factor 
is along $P_i+J_{1i}$ ($\approx e^{-r}$), then $\Delta A$ must be along 
$J_{ij}$ ($\approx e^{-r}$) and $I _{\xi }A _t$ must be along 
$J_{ij}$ (finite), so the whole expression vanishes as $e^{-2r}$.\\
iv. Out of $\Delta A$, $F_t$ and $I _{\xi }A _t$ only $F_t$ has a 
finite contribution at the boundary along $P_1$. That should go together 
with components of the remaining 
factors along the $J_{ij}$ generators, which are either finite or, for $\Delta A$ go to zero as
$\approx e^{-r}$, therefore the whole combination vanishes as $\approx e^{-r}$.\\
The reader may  check the remaining cases and verify that they are all zero except 
for one that is finite:\\
v. $I _{\xi }A _t$ is along $P_i-J_{1i}$ ($\approx e^r$), $\Delta A$ is along $P_i+J_{1i}$ 
($\approx e^{-r}$) and the $F_t$'s are along the $J_{ij}$ generators (finite), yielding a total finite contribution.\\
Using the expression for the charge density eq.(\ref{Noether-charge-diffeo}), the asymptotic expressions given above and the symmetrized trace used we get the explicit expression
\begin{eqnarray}
Q _{\xi} = -2 n \kappa  \int_0^1dt ~\epsilon _{ijk_1l_1...k_{n-1}l_{n-1}}
\Delta \hat{\zeta}^iI _{\xi }\hat{e}^j_{\infty}\nonumber\\
(\hat{R}^{k_1l_1}_{\infty}-  
\hat{\zeta }_t^{k_1}\hat{e}^{l_1}_{\infty}-\hat{e}^{k_1}_{\infty}\hat{\zeta }_t^{l_1})
 ... 
(\hat{R}^{k_{n-1}l_{n-1}}_{\infty}-  
\hat{\zeta }_t^{k_{n-1}}\hat{e}^{l_{n-1}}_{\infty}-\hat{e}^{k_{n-1}}_{\infty}\hat{\zeta }_t^{l_{n-1}})  \label{explicit-Noether-charge-diffeo}
\end{eqnarray}
The previous expression can be simplified, carrying out the integral in the parameter $t$, 
in the particular case of the fixed background ("AdS vacuum") considered at the 
end of the previous subsection.
In that case $\hat{R}^{ij}_{\infty}-  
\hat{\overline{\zeta }}^i\hat{e}^j_{\infty}-\hat{e}^i_{\infty}\hat{\overline{\zeta }}^j=0$, then 
$\hat{R}^{kl}_{\infty}-  
\hat{\zeta }_t^{k}\hat{e}^{l}_{\infty}-\hat{e}^{k}_{\infty}\hat{\zeta }_t^{l}=- t (
\Delta\hat{\zeta }^{k}\hat{e}^{l}_{\infty}+\hat{e}^{k}_{\infty}\Delta\hat{\zeta }^{l})$. Plugging this in eq.(\ref{explicit-Noether-charge-diffeo}) and integrating in $t$ we get
\begin{eqnarray}
Q _{\xi} = (-1)^n   2 \kappa ~\epsilon _{ijk_1l_1...k_{n-1}l_{n-1}}
\Delta \hat{\zeta}^iI _{\xi }\hat{e}^j_{\infty}\nonumber\\
(\Delta\hat{\zeta }^{k_1}\hat{e}^{l_1}_{\infty}+\hat{e}^{k_1}_{\infty}\Delta\hat{\zeta }^{l_1})
 ... 
(\Delta\hat{\zeta }^{k_{n-1}}\hat{e}^{l_{n-1}}_{\infty}+\hat{e}^{k_{n-1}}_{\infty}\Delta\hat{\zeta }^{l_{n-1}})  \label{explicit-Noether-charge-diffeo-background}
\end{eqnarray}
Using again that $\hat{R}^{ij}_{\infty}-  
\hat{\overline{\zeta }}^i\hat{e}^j_{\infty}-\hat{e}^i_{\infty}\hat{\overline{\zeta }}^j=0$ we see that 
$\Delta\hat{\zeta }^{k}\hat{e}^{l}_{\infty}+\hat{e}^{k}_{\infty}\Delta\hat{\zeta }^{l}=-F^{kl}(x,r\rightarrow\infty)
\equiv -F^{kl}_{\infty}$, therefore
\begin{eqnarray}
Q _{\xi} = -  2 \kappa ~\epsilon _{ijk_1l_1...k_{n-1}l_{n-1}}
\Delta \hat{\zeta}^i I _{\xi }\hat{e}^j_{\infty}
F^{k_1 l_1}_{\infty}
 ... 
F^{k_{n-1} l_{n-1}}_{\infty}  \label{explicit-Noether-charge-diffeo-background-1}
\end{eqnarray}
In the previous expressions for $Q _{\xi}$ all the fields are of course taken at $r\rightarrow\infty $, even if 
not specifically denoted (that is for instance the case for $\Delta \hat{\zeta}^i$ in eq.(\ref{explicit-Noether-charge-diffeo-background-1})).

\subsection{Finiteness of the Action } 

In this subsection I discuss the finiteness of the action for field configurations 
with the required asymptotic behaviour, without requiring that those configurations 
should be solutions of the field equations. More precisely, I will show that in the case of a finite
hyper volume boundary \footnote{This is the case for instance if the boundary of the region 
of integration (between two given times) is the Cartesian product of a sphere of dimension $d-2$ times a line interval (the spatial infinity times a the time interval) or in the Euclidean 
case Cartesian product of a sphere of dimension $d-2$ times a circle (the spatial infinity times
a periodic time interval). } the action has not divergences coming from the boundary or the
$r\rightarrow\infty$ part of the integral involved in its definition. 
In holographic AdS-CFT language we may say that there are not ultraviolet divergences. 
We cannot exclude the possibility of divergences coming from singularities 
in the bulk, but for configurations satisfying the field equations ("classical solutions", relevant in saddle point approximations of path integrals) their absence should be a consequence of a kind of 'Cosmic Censorship
Conjecture', implying that any singularity should be hidden behind a horizon that would exclude 
it from the region of integration (at least in the euclidean case). 


 In the case of a boundary with infinite hyper volume (for instance if the boundary is Minkowski space or a de Sitter space) our proof shows that the action per unit boundary volume is finite.  

Finiteness(or finiteness per unit volume) of the euclidean action is necessary for 
instance for finiteness (or finiteness per unit volume) of the thermodynamic 
properties (energy, entropy, temperature, etc.) of black holes.

We need to consider the transgression form, which is to be integrated in a space-time volume between two times (or a period of Euclidean time)
\begin{equation}
\mathcal{T}_{2n+1}\left(  A,\overline{A}\right)  =(n+1)\int_{0}^{1}
dt\ <\Delta{A}F_{t}^{n}>\nonumber
\end{equation}
We observe that all components of $\Delta A$ and $F_t$ are at most finite, so the previous
expression is at most finite. However that is not enough, as it is to be integrated in $r$ to un upper limit of $\infty $. Looking in more detail we see that the only finite component of $\Delta A$ is along $P_i-J_{1i}$, and that must come together with one component of $F_t$ along $P_i+J_{1i}$ ($\approx e^{-r}$) and the remaining components of $F_t$ along $J_{ij}$ (finite). This term goes then as $\approx e^{-r}$, ensuring that when integrated it will not yield boundary divergences. Every other term will have finite $F_t$ components together with $\Delta A$ components that go as $e^{-r}$, as in the previous case. The conclusion is that there will not be divergences coming from the boundary, as far as the integral in the boundary coordinates does not yield an infinite volume (and in that case we have a finite action per unit volume). 

The previous argument actually avoided the subtle point that the Transgression action is defined not by integrating the transgression form in a single bulk manifold but in two (see eq.(\ref{transgression-action})), and it would be valid only in the case in which both manifolds are identical. However we may argue that both bulk manifolds "look the same" towards the boundary and may be identified taking points with the same coordinates $r$ and $x$ (of the kind we have been using) of each manifold as the same, then the cancellations between the $A$ and $\overline{A}$ parts (and the terms containing both)
that ensure the finiteness of the action would hold as before, at least from some finite $r$ to the boundary (which would again show that no divergences come from the boundary region).

Notice that those cancellations do not hold for a pure Chern-Simons Lagrangian density
\begin{equation}
\mathcal{Q}_{2n+1}\left(  A,\overline{A}\right)  =(n+1)\int_{0}^{1}
dt\ <AF_{t}^{n}>
\end{equation}
as $A$ has a component that goes as $e^r$ which yields a finite Lagrangian density and a divergent action.


\section{Action Principle and Boundary Conditions II: Kounterterms}
 
The natural, or rather the naive, vacuum in a field theory corresponds to the configuration in which all fields vanish. One may regard the transgression as a tool to regularize a physical theory, with $A$ being the physical fields and $\overline{A}$ some non dynamical regulator configuration or vacuum to be subtracted. The choice $\overline{A}=0$ (the naive vacuum), which yields Chern-Simons forms, gives infinite values for conserved charges and thermodynamic quantities for black holes as well as an ill-defined action principle. In refs.\cite{motz1,motz3} it was shown that a choice of $\overline{A}$ that properly regularize the action is given by
\begin{equation}
\overline{\omega}^{ij}(x)=\hat{\omega }_{\infty} ^{ij}(x),\;\;\;\overline{\omega}^{1j}=0 ,\;\;\; \overline{e}=0
\end{equation}
I will call this the "Kounterterms vacuum", as it is this choice of $\overline{A}$ that 
leads to the boundary terms associated with the Kounterterms approach to regularization of Chern-Simons and Lovelock theories. Notice that the vielbein does vanish for this configuration, as it does in the naive vacuum, but the spin connection does not. 

We will need
\begin{eqnarray}
\Delta A=e^r\hat{e}^i_{\infty}(P_i-J_{1i})
+\frac{1}{2}\hat{\tau}^i(P_i-J_{1i})+dr~P_1+\nonumber\\
+\frac{1}{2}e ^{-r}[\hat{\gamma}^{ij}+dr\hat{\omega}_r^{ij}]J_{ij}-\frac{1}{2}e ^{-r}\hat{\zeta}^i(P_i+J_{1i})\label{Delta-A-WT1}\\
A_t=t e^r\hat{e}^i_{\infty}(P_i-J_{1i})
+\frac{1}{2}\hat{\omega}_{\infty}^{ij}J_{ij}+\frac{1}{2}t\hat{\tau}^i(P_i-J_{1i})+t~dr~P_1+\nonumber\\
+\frac{1}{2}e ^{-r}t[\hat{\gamma}^{ij}+dr\hat{\omega}_r^{ij}]J_{ij}-\frac{1}{2}e ^{-r}t\hat{\zeta}^i(P_i+J_{1i})\label{A_t-WT1}\\
\delta A_t=t e^r\delta\hat{e}^i_{\infty}(P_i-J_{1i})
+\frac{1}{2}\delta\hat{\omega}_{\infty}^{ij} J_{ij}+\frac{1}{2}t\delta\hat{\tau}^i(P_i-J_{1i})+\nonumber\\
+\frac{1}{2}e ^{-r}t[\delta\hat{\gamma}^{ij}+dr\delta\hat{\omega}_r^{ij}]J_{ij}-\frac{1}{2}e ^{-r}t\delta\hat{\zeta}^i(P_i+J_{1i})\label{delta-A_t-WT1}
\end{eqnarray}
The first two terms of the second member of eq.(\ref{delta-A_t-WT1}) will 
vanish in our discussion of the action principle, as we will require the boundary condition $\delta\hat{e}^i_{\infty} =0$, which in turn implies
$\delta\hat{\omega}_{\infty}^{ij} =0$. We will also need 
\begin{eqnarray}
F_t=\frac{1}{2}F_t^{ij}J_{ij}-t^2[\hat{\zeta } _i\hat{e}^i_{\infty}+\frac{1}{2}e^{-r}\hat{\zeta } _i\hat{\tau}^i]P_1+t(t-1)e^r\hat{e}^i_{\infty}dr(P_i-J_{1i})\nonumber\\
+\{ \frac{t}{2}\hat{D}_{\infty}\hat{\tau } ^i+ 
t^2\hat{\gamma } ^i_{~j}\hat{e}^j_{\infty}+dr[\frac{1}{2}(t\partial _r\hat{\tau } ^i- t^2\hat{\tau } ^i)+t^2\hat{\omega } ^i_{r~j}\hat{e}^j_{\infty} ]\}(P_i-J_{1i})+\nonumber\\
 + \frac{1}{2}e^{-r}t^2\{ \hat{\gamma }^i_{~j}\hat{\tau } ^j+dr \hat{\omega } _{rj}^i\hat{\tau}^j\} (P_i-J_{1i})
-\frac{1}{2}e^{-r}t[\hat{D}_{\infty}\hat{\zeta } ^i
 +dr \partial _r\hat{\zeta }^i]  (P_i+J_{1i})   \nonumber\\
 +\frac{1}{2}e^{-r}t(t-1)\hat{\zeta }^idr  (P_i+J_{1i})
-\frac{1}{2} e^{-2r}t^2\{ \hat{\gamma } ^i_{~j}\hat{\zeta } ^j+dr\hat{\omega } _{rj}^i\hat{\zeta }^j  \}(P_i+J_{1i}) \label{F_t-WT1}
\end{eqnarray}
where
\begin{eqnarray}
F_t^{ij}=\hat{R}^{ij}_{\infty}-  
t^2[\hat{\zeta }^i\hat{e}^j_{\infty}+\hat{e}^i_{\infty}\hat{\zeta }^j]+\nonumber\\
+e^{-r}t[\hat{D}_{\infty}\hat{\gamma } ^{ij}+dr(\partial _r\hat{\gamma }^{ij}-\hat{D}_{\infty}\hat{\omega }_r^{ij})]-\frac{1}{2}e^{-r}t^2(\hat{\zeta }^i\hat{\tau }^j+\hat{\tau }^i\hat{\zeta }^j)
\nonumber\\
+e^{-2r}t^2\{ \hat{\gamma }^i_{~k}\hat{\gamma }^{kj}+ 
dr[\hat{\omega}_{rk}^i\hat{\gamma}^{kj}+\hat{\omega}_{rk}^j\hat{\gamma}^{ik} ]\} 
\label{Fij_t-WT}  
\end{eqnarray}

\subsection{Action Principle}

The action of eq.(\ref{action-trangresion-ads}) yields exactly the same field equations 
for $A$, wheter $\overline{A}$ is taken as fixed or not, but it is important to notice that
the $\overline{A}$ chosen in this section does not 
satisfy the field equations. While $\overline{T}=0$ ensures that the second equation of (\ref{field-equations-transgression-ads}) is satisfied, and $\overline{\omega }^{1i}=0$ implies $\overline{R}^{1i}=0$, which together with $\overline{e}=0$ means that the first equation of (\ref{field-equations-transgression-ads}) is also satisfied, except in the case where the index 1 in the Levi-Civita tensor is the one not summed. In that case 
 $$\epsilon (\overline{R}+\overline{e}^2)^n=\epsilon \overline{R}^n= \overline{E}_n$$ 
 with $\overline{E}_n$ the Euler density of the boundary, which is not necessarily zero.
Therefore the field $\overline{A}$ is to be regarded as a non dynamical field, which is not varied in the 
action principle.  This is fully consistent with our boundary conditions, for which at the boundary 
vielbein is kept fixed $\delta\hat{e}^i_{\infty}=0 $, and therefore the boundary spin connection 
is kept fixed $\delta\hat{\omega}^{ij}_{\infty}=0$, as a consequence of the fact that requiring a finite 
gauge AdS curvature implies a vanishing boundary torsion $\hat{T}^i_{\infty}=0$. 

In the las part of 
this section we will consider a slightly different $\overline{A}$, that only differs from the one 
defined earlier in this section by having $\overline{e}=dr$ instead of just zero. This alternate Kounterterms gauge potential is also taken as fixed $\delta\overline{A}=0$, as its component along $P_1$ and $dr$ is just 1.

We need to verify that the action is an extremum under variations subjected to the 
boundary conditions, if the field equations hold. The first bulk contribution $<F^n\delta A>$ vanishes 
as a result of the field equations for $A$. The second bulk contribution $<\overline{F}^n\delta \overline{A}>$
vanishes because $\delta\overline{A}=0$. We are left with the bulk contribution to the variation, which involves the boundary integral of
$-n(n+1)\int _0^1dt<\Delta AF_t^{n-1}\delta A_t>$. An analysis similar to the one of the previous section shows that in the case at hand there remains in fact a finite contribution at the boundary, coming from $\Delta A$ along 
$P_i-J_{1i}$, $\delta A _t$ along $P_i+J_{1i}$ and the $F_t$'s along $J_{ij}$. The result is proportional to
$$\epsilon _{ijk_1l_1...k_{n-1}l_{n-1}} \hat{e}^i_{\infty}\delta\hat{\zeta}^j_{\infty}F_t^{k_1l_1}...F_t^{k_{n-1}l_{ n-1}}$$
As $F_t^{ij}$ is generically finite, one can check that the vanishing of the previous expression requires
$\delta\hat{\zeta}^j_{\infty}=0$, for its components along the boundary. This condition must be added to the Dirichlet boundary condition $\delta\hat{e}^i_{\infty}=0$  as an independent condition for the Kounterterms action principle.
Together they imply that at the boundary $\delta T^1=0$ (or more precisely its components along the boundary), 
though the reverse is not true.

A particular case in which the condition $\delta\hat{\zeta}^i_{\infty}=0$ would be automatically satisfied
is if the AdS gauge curvature $F$ vanishes at the boundary, because in that case $\hat{\zeta}^i_{\infty}$ is completely 
determined from the boundary vielbein and spin connection (which in turn is determined by the boundary vielbein), as shown by eq.(\ref{C-zeta}). It follows that $\delta\hat{e}^i_{\infty}=0$ implies $\delta\hat{\zeta}^i_{\infty}=0$ in that case.

\subsection{Finite Conserved Charges}

We start with the expression for the diffeomorphism Noether's charge density given by 
eq.(\ref{Noether-charge-diffeo}). Proceeding as in the previous section, we see that the
charge density is indeed finite, with two non vanishing contributions, unlike the case of the 
previous section when we had just one. This contributions come from $\Delta A$ along 
$P_i-J_{1i}$, $I_{\xi } A _t$ along $P_i+J_{1i}$ and the $F_t$'s along $J_{ij}$ on the one hand
and from $\Delta A$ along $P_i+J_{1i}$, $I_{\xi } A _t$ along $P_i-J_{1i}$ and the 
$F_t$'s along $J_{ij}$ on the other hand.

Plugging explicit expressions in eq.(\ref{Noether-charge-diffeo}) we obtain
\begin{eqnarray}
Q_{\xi}=-2 \kappa n\int _0^1dt~t~\epsilon _{ijk_1l_1...k_{n-1}l_{n-1}} [\hat{e}^i_{\infty}I_{\xi}\hat{\zeta}^j_{\infty}- \hat{\zeta}^j_{\infty}I_{\xi}\hat{e}^j_{\infty}]F_t^{k_1l_1}...F_t^{k_{n-1}l_{ n-1}}\label{noether-charge-diffeos-kounterterms}
\end{eqnarray}
where $F_t^{ij}$ only contributes at the boundary through its finite part 
$$(F_t^{ij})_{Finite}=\hat{R}^{ij}_{\infty}-t^2[\hat{\zeta }_{\infty}^i\hat{e}^j_{\infty}+\hat{e}^i_{\infty}\hat{\zeta }_{\infty}^j]$$
It is possible to simplify a little eq.(\ref{noether-charge-diffeos-kounterterms}) by changing the integration parameter from $t$ to $u=t^2$. With any choice of the parameter, the integration can be done by expanding the products and integrating term by term.

A simpler particular case corresponds to having $F$ asymptotically zero, and a constant curvature boundary manifold, for 
which $\hat{R}^{ij}_{\infty}=K \hat{e}^i_{\infty}\hat{e}_{\infty}^j $ where $K$ is a constant. In this
case $\hat{\zeta }_{\infty}^i$ is proportional to $\hat{e}^i_{\infty}$ and $Q_{\xi}$ is reduced to a simple
expression times the boundary volume element.

\subsection{Finiteness of the Action }

In discussing the finiteness of the action a slightly different choice of   
$\overline{A}$ is actually better. Instead of choosing $\overline{e}^ a=0$ we choose
$\overline{e}^1=dr$, that is
\begin{equation}
\overline{\omega}^{ij}(x)=\hat{\omega }_{\infty} ^{ij}(x),\;\;\;\overline{\omega}^{1j}=0 ,\;\;\; \overline{e}^i=0,~~~\overline{e}^ 1=dr
\end{equation}
The difference between the previous choice and this one, for any field of interest, is some term along $dr$. Those terms make no difference as far as the action principle of the conserved charges are concerned, as in both cases what matters are boundary integrals, for which $dr=0$(more precisely, integrals on the boundary of differential forms with any $dr$ in it are zero). However in this subsection we are concerned with forms to be integrated in the bulk, so those differences are relevant.
We now have
\begin{eqnarray}
\Delta A=e^r\hat{e}^i_{\infty}(P_i-J_{1i})
+\frac{1}{2}\hat{\tau}^i(P_i-J_{1i})+\nonumber\\
+\frac{1}{2}e ^{-r}[\hat{\gamma}^{ij}+dr\hat{\omega}_r^{ij}]J_{ij}-\frac{1}{2}e ^{-r}\hat{\zeta}^i(P_i+J_{1i})\label{Delta-A-WT2}\\
A_t=t e^r\hat{e}^i_{\infty}(P_i-J_{1i})
+\frac{1}{2}\hat{\omega}_{\infty}^{ij}J_{ij}+\frac{1}{2}t\hat{\tau}^i(P_i-J_{1i})+dr~P_1+\nonumber\\
+\frac{1}{2}e ^{-r}t[\hat{\gamma}^{ij}+dr\hat{\omega}_r^{ij}]J_{ij}-\frac{1}{2}e ^{-r}t\hat{\zeta}^i(P_i+J_{1i})\label{A_t-WT2} 
\end{eqnarray}
while $\delta A_t$ is just as before given by eq.(\ref{delta-A_t-WT1})
For $F_t$ we now have 
\begin{eqnarray}
F_t=\frac{1}{2}F_t^{ij}J_{ij}-t^2[\hat{\zeta } _i\hat{e}^i_{\infty}+\frac{1}{2}e^{-r}\hat{\zeta } _i\hat{\tau}^i]P_1+\nonumber\\
+\{ \frac{t}{2}\hat{D}_{\infty}\hat{\tau } ^i+ 
t^2\hat{\gamma } ^i_{~j}\hat{e}^j_{\infty}+dr[\frac{t}{2}(\partial _r\hat{\tau } ^i- \hat{\tau } ^i)+t^2\hat{\omega } ^i_{r~j}\hat{e}^j_{\infty} ]\}(P_i-J_{1i})+\nonumber\\
 + \frac{1}{2}e^{-r}t^2\{ \hat{\gamma }^i_{~j}\hat{\tau } ^j+dr \hat{\omega } _{rj}^i\hat{\tau}^j\} (P_i-J_{1i})
-\frac{1}{2}e^{-r}t[\hat{D}_{\infty}\hat{\zeta } ^i
 +dr \partial _r\hat{\zeta }^i]  (P_i+J_{1i})-\nonumber\\
-\frac{1}{2} e^{-2r}t^2\{ \hat{\gamma } ^i_{~j}\hat{\zeta } ^j+dr\hat{\omega } _{rj}^i\hat{\zeta }^j  \}(P_i+J_{1i}) \label{F_t-WT2}
\end{eqnarray}
where $F_t^{ij}$ is just the same of eq.(\ref{F_t-WT1}). Notice that there are no divergent components in this form of $F_t$.

As at the end of the previous section, we want to show that no divergence comes from the $r\rightarrow\infty$ part
of the integral that defines the action. What we need is the asymptotic behavior of $<\Delta{A}F_{t}^{n}>$ be
such that when it is integrated in $r$ would not yield a divergence. Proceeding as in the previous section we see that every possible term either vanishes because of the trace or goes to zero as $e^{-r}$ or faster, except for one finite term. That finite term would give a divergence linear on $r$ when integrated, which would be a logarithmic divergence in the standard Fefferman-Graham radial coordinate $\rho$. The finite term corresponds to
$\Delta A$ along $P_i-J_{1i}$, one $F_t$ along $P_i+J_{1i}$ and the rest of the $F_t$'s
 along $J_{ij}$. It is explicitly given by
\begin{eqnarray}
-2\kappa n\int _0^1dt~t~\epsilon _{ijk_1l_1...k_{n-1}l_{n-1}} \hat{e}^i_{\infty}                            [\hat{D}_{\infty}\hat{\zeta } ^j
 +dr \partial _r\hat{\zeta }^j]F_t^{k_1l_1}...F_t^{k_{n-1}l_{ n-1}} 
\end{eqnarray}
It follows that in order to avoid divergences coming from $r\rightarrow\infty$ we need to require that
$[\hat{D}_{\infty}\hat{\zeta } ^i+dr \partial _r\hat{\zeta }^i]$ goes to zero as a function of $r$ fast 
enough to make the integral in $r$ finite or zero for $r\rightarrow\infty$. Our assumptions to this point would only imply that this expression is finite, so this is an additional condition on $\hat{\zeta } ^i(x,r)$.

Near the boundary $\hat{\zeta } ^i(x,r)=\hat{\zeta } ^i_{\infty}(x)+e^{-r}f ^i(x,r)$, where $f ^i(x,r)$ 
could even be divergent, but only in such a way that the combination $e^{-r}f ^i(x,r)$ goes to zero. 
We have $dr \partial _r\hat{\zeta }^j=dr \partial _r[e^{-r}f ^i(x,r)]$, but the integral of this form, 
times some finite things,  
as $\int ^{\infty}e^{-r}f ^i(x,r)$ \footnote{If we consider just $\int ^{\infty}dr \partial _r[e^{-r} f^i(x,r)]$ the integral would vanish as $e^{-r}f ^i(x,r)$, but $\int ^{\infty}dr \partial _r[e^{-r} f^i(x,r)][finite~part]$ would go as 
$\approx\int ^{\infty}dr~e^{-r}f ^i(x,r)$ asymptotically, as it may be seen integrating by parts.}. 

We also have 
$\hat{D}_{\infty}\hat{\zeta } ^i(x,r)=\hat{D}_{\infty}\hat{\zeta }_{\infty} ^i (x)+e^{-r}\hat{D}_{\infty}f ^i(x,r)$. 
Here $\hat{D}_{\infty}\hat{\zeta }_{\infty} ^i (x)$ is in principle finite, while $e^{-r}\hat{D}_{\infty}f ^i(x,r)$
goes asymptotically as $e^{-r}f ^i(x,r)$, because the covariant derivative $\hat{D}_{\infty}$ does 
not change the behavior on $r$. The bulk integrals of these terms must be zero or finite.

All the previous considerations on the additional conditions on $\hat{\zeta } ^i(x,r)$ for a finite action boil down to:\\
i.  $\hat{D}_{\infty}\hat{\zeta }_{\infty} ^i(x)=0$ and \\
ii. $\int ^{\infty}dr~e^{-r}f ^i(x,r)$ must be finite or zero.

Condition i. is satisfied automatically in the particular case that the AdS gauge curvature vanishes asymptotically and 
the boundary is a constant 
curvature manifold, as in that case
$\hat{\zeta }_{\infty} ^i (x)$ is proportional to the boundary vielbein. The second condition seems rather weak,
as it only exclude rather strange configurations, having for instance $f ^i\approx e^r/r $ (which probably are excluded from solutions to the field equations).

\section{Examples of configurations in radially simple coordinates}
 
 \subsection{Spaces with vanishing asymptotic gauge curvature and boundaries of constant curvature}

In general, in the basis of the boundary vielbein
\begin{eqnarray}
\hat{R}^{ij}_{\infty}=\frac{1}{2}\mathcal{R}^{ij}_{~~kl}\hat{e}^k_{\infty}\hat{e}^l_{\infty}
\end{eqnarray}
For a constant curvature boundary $\hat{R}^{ij}_{\infty}=K\hat{e}^k_{\infty}\hat{e}^l_{\infty}$ where 
$K$ is a constant. Then
\begin{eqnarray}
\mathcal{R}^{ij}_{~~kl}=K[\delta ^i _k\delta ^j_l-\delta ^i_l\delta ^j_k]~~,~~
\mathcal{R}^{i}_{~k}=K(D-1)\delta ^i _k~~,~~\mathcal{R}=K(D-1)D
\end{eqnarray}
If $F=0$ we can use eq.(\ref{C-zeta}), which yields $C^i_{~k}=\frac{K}{2}\delta ^i_k$, implying
\begin{eqnarray}
\hat{\zeta}^i_{\infty}(x)=\frac{K}{2}\hat{e}^i_{\infty}(x)
\end{eqnarray}
If $\hat{\tau}^i$, $\hat{\gamma}^{ij}$ and $\hat{\omega}_r^{ij}$ vanish at the boundary, then the gauge 
curvature $F$ vanishes at the boundary.

\subsection{Black Holes in arbitrary dimension}

The black hole solutions discussed in this section were introduced and studied elsewhere (
see \cite {dimensionally,topo1,topo2,scan} and references therein). Here we just recast those solutions in radially simple coordinates and compute conserved quantities with the formulas given above.
There are many black hole solutions to different gravity theories with a line element of the
generic form
\begin{eqnarray}
 ds^2=-\Delta ^2 (\mathfrak{r})dt^2+\frac{d\mathfrak{r}^2}{\Delta ^2 (\mathfrak{r})}+
 \mathfrak{r}^2d\Sigma ^2
\end{eqnarray}
where $\Delta (\mathfrak{r})$ is some function of the "radial" coordinate $\mathfrak{r}$, and 
$d\Sigma ^2$ spatial boundary line element. The boundary may be a sphere, or a different constant curvature 
manifold,
even a non compact one (in that case we have a "Black Brane"), and it has coordinates 
$x^{\underline{I}}$, where the underlined upper case (space-time) latin index $\underline{I}$ 
takes all the allowed 
values except by 0 and 1.
The vielbein and spin connection associated to this metric are, assuming zero torsion
\begin{eqnarray}
 e^0=\Delta (\mathfrak{r})dt~~,~~e^1=\frac{d\mathfrak{r}}{\Delta (\mathfrak{r})}~~,~~e^{\underline{i}}=
 \mathfrak{r}\hat{e}^{\underline{i}}_{\infty}(x)\label{vielbein-BlackHole}\\
 \omega ^{01}=\Delta 'e^0~~,~~\omega ^{0\underline{i}}=0~~,
~~\omega ^{1\underline{i}}=-\Delta \hat{e}^{\underline{i}}_{\infty}~~,
~~\omega ^{\underline{i}\underline{j}}=\hat{\omega } ^{\underline{i}\underline{j}}_{\infty}(x)\label{spin-connection-BlackHole}
\end{eqnarray}
where again underlined lower case (tangent) latin indices take all the allowed values except by 0 
and 1, and primes mean derivatives with respect to $\mathfrak{r}$.
The curvature two-form has components
\begin{eqnarray}
 R^{0\underline{i}}=-\left[\frac{\Delta ^2}{2}\right]'e^0\hat{e}^{\underline{i}}_{\infty}~~,~~
R^{01}=-\left[\frac{\Delta ^2}{2}\right]'' e^0e^1 \nonumber\\
R^{\underline{i}\underline{j}}=\hat{R}_{\infty}^{\underline{i}\underline{j}}-\Delta ^2 \hat{e}^{\underline{i}}_{\infty}\hat{e}^{\underline{j}}_{\infty}~~,~~R^{1\underline{i}}=-\left[\frac{\Delta ^2}{2}\right]'e^1\hat{e}^{\underline{i}}_{\infty}
\end{eqnarray}
The corresponding AdS gauge curvatures are then
\begin{eqnarray}
 F^{0\underline{i}}=\left[\mathfrak{r}-\left(\frac{\Delta ^2}{2}\right) '\right] e^0\hat{e}^{\underline{i}}_{\infty}~~,~~
F^{01}=\left[1-\left( \frac{\Delta ^2}{2}\right) ''\right] e^0e^1 \nonumber\\
F^{\underline{i}\underline{j}}=\hat{R}_{\infty}^{\underline{i}\underline{j}}+[\mathfrak{r}^2-\Delta ^2] \hat{e}^{\underline{i}}_{\infty}\hat{e}^{\underline{j}}_{\infty}~~,~~F^{1\underline{i}}=\left[\mathfrak{r}-\left( \frac{\Delta ^2}{2}\right)'\right]e^1\hat{e}^{\underline{i}}_{\infty}
\end{eqnarray}
An important particular case corresponds to AdS space-time, for which 
\begin{eqnarray}
\Delta ^{(AdS)}(\mathfrak{r})=\sqrt{\mathfrak{r}^2 +1} 
\end{eqnarray}
and the boundary manifold is a constant curvature sphere, so that 
$\hat{R}_{\infty}^{\underline{i}\underline{j}}=\hat{e}^{\underline{i}}_{\infty}\hat{e}^{\underline{j}}_{\infty}$.
One can check that in this case $F=0$. Notice that for this AdS metric $\Delta (\mathfrak{r})\approx \mathfrak{r}$ 
when $\mathfrak{r}\rightarrow\infty$, which is actually a generic property for black hole solutions in theories
with asymptotically AdS behavior.

In order to pass to radially simple coordinates we just keep the same $t$ and $x^{\underline{I}}$ coordinates, 
and change the radial coordinate in such a way that $dr^2=\frac{d\mathfrak{r}^2}{\Delta ^2 (\mathfrak{r})}$ or
$dr =\pm \frac{d\mathfrak{r}}{\Delta (\mathfrak{r})}$. We choose the plus sign in order 
to have $r\rightarrow\infty$ when $\mathfrak{r}\rightarrow\infty$. Integrating we get
\begin{eqnarray}
r =\int \frac{d\mathfrak{r}}{\Delta (\mathfrak{r})} \label{r-solodukhin-r-standard}
\end{eqnarray}
There is an undetermined integration constant in the previous expression, which we will choose below. 
Notice that if $\Delta (\mathfrak{r})\approx \mathfrak{r}$ when $\mathfrak{r}\rightarrow\infty$, as said above, then
$\mathfrak{r}\approx e^r$ when $r\rightarrow\infty$, as we may expect. The freedom to choose the integration constant is used to make the proportionality constant in the asymptotic expression $\mathfrak{r}\approx e^r$ equal to 1, in order to recover the standard form of the metric in radially simple coordinates given in the text.
The explicit expressions for the relevant fields, for the black hole solution, are
\begin{eqnarray}
e^1=dr ~~,~~e^0 =\Delta ( \mathfrak{r}) dt= \Delta ( \mathfrak{r})\hat{e}^0_{\infty}
~~,~~e^{\underline{i}}=\mathfrak{r}\hat{e}^{\underline{i}}_{\infty}(x)\nonumber \\
 \omega ^{01}=\left(\frac{\Delta ^2( \mathfrak{r})}{2}\right) ' dt= \left(\frac{\Delta ^2( \mathfrak{r})}{2}\right) ' \hat{e}^0_{\infty}~~,~~\omega ^{0\underline{i}}=0 \nonumber\\
\omega ^{1\underline{i}}=-\Delta ( \mathfrak{r}) \hat{e}^{\underline{i}}_{\infty}(x)~~,
~~\omega ^{\underline{i}\underline{j}}=\hat{\omega } ^{\underline{i}\underline{j}}_{\infty}(x)
\end{eqnarray}
where we are regarding $\mathfrak{r}$ as a function of $r$, from eq.(\ref{r-solodukhin-r-standard}) with a 
suitable integration constant. These means that
\begin{eqnarray}
\zeta ^0 =e^{-r}\left[\left(\frac{\Delta ^2( \mathfrak{r})}{2}\right) '- \Delta ( \mathfrak{r}) \right] dt~~,~~
\zeta ^{\underline{i}} =e^{-r}\left[\Delta ( \mathfrak{r})- \mathfrak{r}\right] \hat{e}^{\underline{i}}_{\infty}\nonumber\\
\tau ^0 =\left\{ e^{-r}\left[\left(\frac{\Delta ^2( \mathfrak{r})}{2}\right) '+ \Delta ( \mathfrak{r})\right]-2 \right\} dt
\nonumber\\
\tau ^{\underline{i}} =
\left\{ e^{-r}\left[\Delta ( \mathfrak{r})+ \mathfrak{r}\right]-2\right\} \hat{e}^{\underline{i}}_{\infty}~~,~~
\gamma ^{ij}=0~~,~~\omega ^{ij}_r=0 \label{blackhole-zeta-tau-gamma}
\end{eqnarray}
If indeed $\Delta ( \mathfrak{r})\approx \mathfrak{r}\approx e^r$ when $r\rightarrow\infty$, then the leading terms in the previous equations cancel, as expected.

\subsection{Chern-Simons Black Holes}

In the concrete case of Black Holes in Chern-Simons Gravity the function $\Delta(\mathfrak{r})$ is
\begin{eqnarray}
\Delta(\mathfrak{r})=\sqrt{\mathfrak{r}^2-\sigma +1} 
\end{eqnarray}
where the constant $\sigma$ is given as $\sigma =(2G~m+1)^{\frac{1}{n}}$ in terms of the 
"Newton constant" $G$  of the theory and the mass $m$ of the black hole. The constant $\kappa$ that appears in the invariant trace
and therefore in the action is related to $G$ by $\kappa =\frac{1}{2G(d-2)!\Omega _{d-2}}$, where
$\Omega _{d-2}$ is the volume of a $d-2$ dimensional sphere of unit radius.

With this $\Delta (\mathfrak{r})$ the AdS gauge curvature are explicitly
\begin{eqnarray}
 F^{0\underline{i}}=0~~,~~
F^{01}=0\nonumber\\
F^{\underline{i}\underline{j}}=\hat{R}_{\infty}^{\underline{i}\underline{j}}+[\sigma -1] \hat{e}^{\underline{i}}_{\infty}\hat{e}^{\underline{j}}_{\infty}~~,~~F^{1\underline{i}}=0
\end{eqnarray}
As the torsion is zero, then $F^a=0$. It is straightforward to check that the field equations are satisfied.

The explicit form of the coordinate change resulting from eq.(\ref{r-solodukhin-r-standard}) depends
on whether $\sigma >1$, $\sigma <1$ or $\sigma =1$. If $\sigma <1$ then
\begin{eqnarray}
r=\sinh ^{-1}\left( \frac{\mathfrak{r}}{\sqrt{1-\sigma}}\right)+ \frac{1}{2}\ln \left(\frac{1-\sigma}{4}\right)\\
\mathfrak{r}= \sqrt{1-\sigma} \sinh \left[ r-\frac{1}{2}\ln \left(\frac{1-\sigma}{4}\right)\right]
\end{eqnarray}
while if $\sigma >1$ then
\begin{eqnarray}
r=\cosh ^{-1}\left( \frac{\mathfrak{r}}{\sqrt{\sigma -1}}\right)+ \frac{1}{2}\ln \left(\frac{\sigma -1}{4}\right)\\
\mathfrak{r}= \sqrt{\sigma -1} \cosh \left[ r-\frac{1}{2}\ln \left(\frac{\sigma -1}{4}\right)\right]
\end{eqnarray}
The constants of integration were chosen to make $\mathfrak{r}\approx e^r$ with proportionality constant equal to 1 when $r\rightarrow\infty$. When $\sigma =1$ we get
\begin{eqnarray}
r=\ln \left( \mathfrak{r} \right)~~,~~\mathfrak{r}= e^r
\end{eqnarray}
where the integration constant was chosen to be zero. It clear that the asymptotic behavior 
of the components of the gauge potential is the expected considering that 
the gauge curvature is indeed finite. That can be explicitly seen from the previous 
expressions and eq.(\ref{blackhole-zeta-tau-gamma}). In particular, we will use in the evaluation of 
conserved charges that
\begin{eqnarray}
\hat{\zeta } ^0 _{\infty}=-\frac{(1-\sigma )}{2}\hat{e}^0_{\infty }=-\frac{ (1-\sigma )}{2}dt~~,~~
\hat{\zeta }^{\underline{i}} _{\infty}= \frac{(1-\sigma )}{2}\hat{e}^{\underline{i}}_{\infty}
\end{eqnarray}

\subsection{Rotating BTZ Black Hole in 2+1 dimensions}

The Chern-Simons rotating BTZ black hole solution in 2+1 dimensions has some charateristics that 
are more generic than the ones of the higher dimensional black holes of the previous subsection.
In this case we have
\begin{eqnarray}
 e^0=\Delta (\mathfrak{r})dt~~,~~e^1=\frac{d\mathfrak{r}}{\Delta (\mathfrak{r})}
~~,~~e^{2}=\mathfrak{r}d\phi -\frac{J}{2\mathfrak{r}}dt\label{vielbein-2+1-BTZ}\\
 \omega ^{01}=\mathfrak{r}dt -\frac{J}{2\mathfrak{r}}d\phi ~~,
~~\omega ^{12}=-\Delta d\phi~~,
~~\omega ^{02}= -\frac{J}{2\mathfrak{r}^2\Delta}d\mathfrak{r}\label{spin-connection-2+1-BTZ}
\end{eqnarray}
where $J$ is a constant ("angular moment") and
\begin{eqnarray}
\Delta(\mathfrak{r})=\sqrt{\mathfrak{r}^2-M +\frac{J^2}{4\mathfrak{r}^2}} 
\end{eqnarray}
with $M$ a constant ("mass"). The AdS gauge curvature corresponding to this gauge potential
is zero, as it should to be a solution of the field equations. The change of coordinates given by 
eq.(\ref{r-solodukhin-r-standard}) depends now in whether $J^2-M^2 >0$, 
$J^2-M^2 < 0$ or $J^2-M^2 =0$. If $J^2-M^2 >0$ then
\begin{eqnarray}
r=\frac{1}{2}\sinh ^{-1}\left( \frac{2\mathfrak{r}^2-M}{\sqrt{J^2-M^2}}\right)+ \frac{1}{2}\ln \left(\frac{\sqrt{J^2-M^2} }{4}\right)\\
\mathfrak{r}=  \sqrt{\frac{\sqrt{J^2-M^2}}{2}\sinh \left[ 2r-\frac{1}{2}\ln \left(\frac{\sqrt{J^2-M^2} }{4}\right)\right]
+\frac{M}{2}}
\end{eqnarray}
If $J^2-M^2 <0$ then
\begin{eqnarray}
r=\frac{1}{2}\cosh ^{-1}\left( \frac{2\mathfrak{r}^2-M}{\sqrt{M^2-J^2}}\right)+ \frac{1}{2}\ln \left(\frac{\sqrt{M^2-J^2} }{4}\right)\\
\mathfrak{r}=  \sqrt{\frac{\sqrt{M^2-J^2}}{2}\cosh \left[ 2r-\frac{1}{2}\ln \left(\frac{\sqrt{M^2-J^2} }{4}\right)\right]
+\frac{M}{2}}
\end{eqnarray}
If $J^2-M^2 =0$ then
\begin{eqnarray}
r=\frac{1}{2}\ln\left( \frac{2\mathfrak{r}^2-M}{2}\right) \\
\mathfrak{r}=  \sqrt{ e^{2r}
+\frac{M}{2}}
\end{eqnarray}
By getting a factor $e^r$ out of the second line of any of the previous 
expressions and expanding what remains 
in powers of $e^{-r}$ we can see that the gauge potential does indeed have 
the expected kind of asymptotic behavior.

\subsection{Some calculations of Noether's charges}

The black hole Noether's charges have been computed elsewhere (see \cite{motz3} and references therein),
here we just check that the expression for Noether's charge densities found agree with known results, as they should. 

\subsubsection{Mass of Chern-Simons black holes in arbitrary dimension: Backgrounds}

The Noether charge corresponding to $\xi =\frac{\partial ~}{\partial t}$ is the total energy (or mass) $E$ of the black hole configuration, in then case of the spacial boundary being a sphere 
(in the case of a black brane it is the energy by unit volume of the boundary). The actual conserved mass comes from integrating $Q_{\xi}$ at the boundary  $\partial\Sigma $
of a constant time slice  $\Sigma $ ("space" at that time) of the space-time manifold $\mathcal{M}$. Neither $dr$ nor $dt$ have support on 
$\partial\Sigma$.

i. In the case in which $\overline{A}$ is the fixed AdS vacuum, which corresponds 
to taking $\overline{\sigma}=0$. We had $\hat{\zeta }^{\underline{i}} _{\infty}= \frac{(1-\sigma )}{2}\hat{e}^{\underline{i}}_{\infty}$, then 
$\Delta\hat{\zeta }^{\underline{i}} _{\infty}= -\frac{\sigma }{2}\hat{e}^{\underline{i}}_{\infty}$, 
and $I _{\xi }\hat{e}^j_{\infty}=0$ unless $j=0$,
in which case $I _{\xi }\hat{e}^0_{\infty}=1$. Using eq.(\ref{explicit-Noether-charge-diffeo-background})
we get \footnote{There is a minus sign coming from bringing  the index $0$ to the first place in the Levi-Civita tensor. Another minus sign comes from the fact that the canonical volume element of
the space-time manifold is $dtdre^{1}...e^{d-2}$, then the induced volume element in the boundary consistent with Stokes theorem is $-dt\hat{e}^{1}_{\infty}...\hat{e}^{d-2}_{\infty}$ (from interchanging $dt$ and $dr$ before integrating on $r$), then the volume element on $\partial\Sigma$ is
$-\hat{e}^{1}_{\infty}...\hat{e}^{d-2}_{\infty}$. The product of both minus signs gives 1.}
\begin{eqnarray}
E=\int _{\partial\Sigma}Q _{\xi} = \kappa ~\sigma ^n\epsilon _{\underline{i}_1...\underline{i}_{d-2}}
\int _{\partial\Sigma}\hat{e}^{\underline{i}_1}_{\infty}...\hat{e}^{\underline{i}_{d-2} }_{\infty} =
\kappa ~\sigma ^n(d-2)!\Omega _{d-2}
\end{eqnarray}
where $\Omega _{d-2}$ is the volume of the $d-2$ sphere.  With the expressions given before $\sigma =(2G~m+1)^{\frac{1}{n}}$ and  $\kappa =\frac{1}{2G(d-2)!\Omega _{d-2}}$ we get
\begin{eqnarray}
E= m+\frac{1}{2G}=m-m_{AdS}
\end{eqnarray}
where we defined the energy of AdS space-time by $m_{AdS}\equiv -\frac{1}{2G}$.

ii. If $\overline{A}$ is taken to be another black hole solution, proceeding as before, using 
eq.(\ref{explicit-Noether-charge-diffeo}) and performing a simple integral in the parameter we get
\begin{eqnarray}
E= \int _{\partial\Sigma}Q _{\xi}=m-\overline{m}
\end{eqnarray}
as expected and in agreement with ref.\cite{motz3}.

\subsubsection{Mass of Chern-Simons black holes in arbitrary dimension: Kounterterms}

The mass of chern-Simons black holes, in the Kounterterms case, can be computed 
using eq.(\ref{noether-charge-diffeos-kounterterms}). Proceeding as we did with the backgrounds calculation,
and performing a simple integral in the interpolating parameter, we get
\begin{eqnarray}
E=\int _{\partial\Sigma}Q _{\xi} = \kappa ~\left[ \sigma ^n -1\right] (d-2)!\Omega _{d-2}=m
\end{eqnarray} 
Notice that in the backgrounds case we always get the difference of values that can be associated to each configuration.
The fact that in the kounterterms case we just get $m$ can be taken as a justification to consider the $\overline{A}$ corresponding to the kounterterms as the true vacuum of the theory.

\subsubsection{Mass and angular momentum of rotating BTZ black hole}

For the BTZ rotating black hole in 2+1 dimensions we get
\begin{eqnarray}
\hat{\zeta}^0_{\infty}=\frac{M}{2}dt-\frac{J}{2}d\phi ~~,~~\hat{e}^0_{\infty}=dt\\
\hat{\zeta}^2_{\infty}=-\frac{M}{2}d\phi+\frac{J}{2}dt ~~,~~\hat{e}^2_{\infty}=d\phi
\end{eqnarray}
The BTZ solution with $M=-1$ and $J=0$ is just the AdS vacuum. Notice that only $d\phi $ has support in the spatial boundary.

The energy (mass) corresponding to the Noether's charge for $\xi =\frac{\partial ~}{\partial t}$ in the case of backgrounds and with $\overline{A}$ given by the AdS vacuum is, using eq.(\ref{explicit-Noether-charge-diffeo-background})
\begin{eqnarray}
E= \frac{M}{2G}+\frac{1}{2G}=m-m_{AdS}
\end{eqnarray}
where we defined the energy (mass) of the black hole by $m=\frac{M}{2G}$ and the energy of AdS space-time by $m_{AdS}\equiv -\frac{1}{2G}$.

The angular momentum corresponding to the Noether's charge for $\xi =\frac{\partial ~}{\partial \phi}$ in the case 
of backgrounds and with $\overline{A}$ given by the AdS vacuum is, using eq.(\ref{explicit-Noether-charge-diffeo-background})
\begin{eqnarray}
\mathfrak{J}= \frac{J}{2G}
\end{eqnarray}
which we interpret as meaning that the angular momentum of the black hole is $\mathfrak{J}$ and 
the angular momentum of the AdS is zero.

\section{Discussion and Comments}

An important issue that we intend to investigate concerns what are the 
boundary data (allowing a finite asymptotic curvature) required to reconstruct the bulk configuration 
when the field equations hold,
in the sense of refs.\cite{deharo-skenderis-solodukhin, banados-miskovic-theisen,banados-olea-theisen}, 
as well as detailed understanding of the structure 
of the asymptotic expansion in that case.
While the main aim of this work is not tha AdS-CFT conjecture \cite{maldacena,witten-ads,Gubser}, the previous point would be relevant for a holographic interpretation of CS-AdS gravity as dual of a boundary CFT.

It would be interesting to understand the problems discussed in the present work for
the case of the so called "exotic Chern-Simons actions", for which the gauge group
is also the AdS group but the invariant tensor used in the construction of the action is not
the Levi-Civita tensor but a combination of standard traces 
(see for instance \cite{zanelli-lectures} and references therein).

The kind of expansion considered here may be of interest in the context of Lovelock gravities
\cite{lovelock-1,lovelock-2,zwiebach,zumino} and their holographic 
interpretation. Those theories have been studied mostly in their metric/torsion-free formulation,
but their natural setting is a firt order formulation with independent vielbein and spin connection.
Furthermore, it has been shown that the boundary terms coming from Transgressions are well 
suited to regularize those theories, both in the particular case of General Relativity \cite{motz2,transgression-gr}
and in the generic case \cite{olea-kounterterms,kofinas-olea,Olea-Miskovic-2}.

Having a single action with a doubling of the field content, as it is naturally the case by construction
in the Transgressions,
with one of the fields (which may be chosen as a "vacuum") regulating the other, 
may be suggestive of a wider conceptual framework where a dynamical mechanism that
introduce scales in an originally scale free theory is built in from the start.
This sort of mechanism, with the second field introduced in a  more or less ad hoc fashion 
has been used
in several attempts to solve the cosmological constant problem in the last decades, and appears as 
an essential ingredient of the recent trend of Double Field Theory (see for instance \cite{aldazabal}). 
This kind of theoretical structure, where the action does contain its own regulator in a dynamical fashion, may
be seen as alternative to the Effective Field Theory approach, and it may be somewhat
related to dimensional transmutation ideas.

{\bf Acknowledgments:} I am grateful to O. Miskovic and R. Olea for discussions and comments.
I had financial support from the {\it Sistema Nacional de Investigadores} (SNI), of the {\it Agencia Nacional de Investigaci\'on e Innovaci\'on} (ANII) of the {\it Rep\'ublica Oriental del  Uruguay} while most of the  work presented 
here was under way. \newline

\appendix

\end{document}